\documentclass[preprint,12pt,authoryear]{elsarticle}

\usepackage{amssymb}    
\usepackage{amsthm}     
\usepackage{amsmath}    
\usepackage{lineno}     
\usepackage{booktabs}   
\usepackage{rotating}   
\usepackage{subfigure}  
\usepackage{subcaption}
\usepackage{adjustbox}  
\usepackage{booktabs}   
\usepackage{multirow}   
\usepackage[dvipsnames]{xcolor}         
\usepackage{threeparttable}             
\usepackage[margin=0.85in]{geometry}    
\usepackage[skip=0.6\baselineskip]{caption} 
\usepackage{array}      
\newcolumntype{L}[1]{>{\raggedright\let\newline\\\arraybackslash\hspace{0pt}}m{#1}}
\newcolumntype{C}[1]{>{\centering\let\newline\\\arraybackslash\hspace{0pt}}m{#1}}
\newtheorem{prop}{Proposition}          
\newtheorem{assumption}{Assumption}     
\newtheorem{observation}{Observation}   
\usepackage{lipsum}                     
\usepackage{epsfig}                     
\usepackage[makeroom]{cancel}           

\usepackage{hyperref}                   
\hypersetup{colorlinks=true}

\journal{\textit{Transportation Research Part B: Methodological}. Please cite: \href{https://doi.org/10.1016/j.trb.2025.103386}{10.1016/j.trb.2025.103386}}

\begin{document}

\begin{frontmatter}

\title{The fragile nature of road transportation networks}

\author[a]{Linghang Sun\corref{cor1}} \ead{lisun@ethz.ch}
\author[a,b]{Yifan Zhang}
\author[c]{Cristian Axenie}
\author[d]{Margherita Grossi}
\author[a]{Anastasios Kouvelas}
\author[a]{Michail A. Makridis}

\affiliation[a]{organization={Institute for Transport Planning and Systems, ETH Zürich},
            city={Zurich},
            postcode={8093}, 
            country={Switzerland}}
\affiliation[b]{organization={Department of Computer Science, City University of Hong Kong (Dongguan)},
            city={Dongguan, Guangdong},
            postcode={523808}, 
            country={China}}
\affiliation[c]{organization={Computer Science Department and Centre for Artificial Intelligence, Technische Hochschule Nürnberg},
            city={Nuremberg},
            postcode={90489}, 
            country={Germany}}
\affiliation[d]{organization={Intelligent Cloud Technologies Lab, Huawei Munich Research Center},
            city={Munich},
            postcode={80992}, 
            country={Germany}}

\cortext[cor1]{Corresponding author.}

\begin{abstract}
Major cities worldwide experience problems with the performance of their road transportation networks, and the continuous increase in traffic demand presents a substantial challenge to the optimal operation of urban road networks and the efficiency of traffic control strategies. The operation of transportation systems is widely considered to display fragile property, i.e., the loss in performance increases exponentially with the linearly growing magnitude of disruptions. Meanwhile, the risk engineering community is embracing the novel concept of antifragility, enabling systems to learn from past events and exhibit improved performance under disruptions of previously unseen magnitudes. In this study, based on established traffic flow theory knowledge, namely the Macroscopic Fundamental Diagram (MFD), we first conduct a rigorous mathematical analysis to theoretically prove the fragile nature of road transportation networks. Subsequently, we propose a skewness-based indicator that can be readily applied to cross-compare the degree of fragility for different networks solely dependent on the MFD-related parameters. Finally, we implement a numerical simulation calibrated with real-world network data to bridge the gap between the theoretical proof and the practical operations, with results showing the reinforcing effect of higher-order statistics and stochasticity on the fragility of the networks. This work aims to demonstrate the fragile nature of road transportation networks and guide researchers towards adopting the methods of antifragile design for future networks and traffic control strategies.
 
\end{abstract}



\end{frontmatter}


\section{Introduction}
\label{sec: introduction}

As reported by both the \cite{us_department_of_transportation_national_2019} and the \cite{federal_statistical_office_of_switzerland_mobilitat_2020}, motorized road traffic before the pandemic has experienced an approximate $50\%$ growth over the past few decades, and multiple studies have confirmed the recovery of motorized traffic from this public health emergency \citep{buchel_covid-19_2022, marra_impact_2022}. Researchers have also found that the continuous growth in traffic volume has consequently contributed to a rise in disruptions, such as severe congestion and more frequent accidents \citep{albalate_relationship_2021}. Although demand management \citep{yildirimoglu_demand_2020}, public transportation \citep{ouyang_continuum_2014}, and road pricing \citep{genser_dynamic_2022} mark the evolution of modern transportation management, it is expected that this upward trend will continue in the coming decades, even under the political and behavioral shifts \citep{matthias_modelling_2020}, such as promoting bicycle-friendly design \citep{ni_bicycle_2024} and working from home \citep{zhang_long-term_2021}. Another emerging transformation that shall not be disregarded is the deployment of Autonomous Vehicles (AVs) and autonomous mobility-on-demand, which may account for more than $10\%$ growth of induced demand \citep{nahmias-biran_who_2021}. Therefore, future road networks are expected to experience a further increase in traffic demand.

Meanwhile, there is a common understanding that road transportation networks can exhibit fragile properties. Fragility signifies a system's susceptibility to exponentially escalating performance deterioration as disruptions increase in their magnitude. One well-known and intuitive example of such fragile characteristics is the BPR function \citep{us1964traffic} and its variations, which have been extensively applied in estimating link or route travel time \citep{lo_degradable_2006, ng_computationally_2010}. The original BPR function can be formulated as $T = T_{ff} \left( 1+0.15 \left( {q}/{q_{\rm max}} \right)^4 \right)$ and illustrates with empirical data that at the link level, as the traffic flow $q$ approaches the capacity of a link $q_{\rm max}$, the travel time $T$ is revealed to grow exponentially compared to the free flow travel time $T_{ff}$. Note that the BPR function may not be perfectly realistic, and its coefficients need recalibration for a given region due to distinct traffic environments and driving behavior \citep{spiess_technical_1990, suh_highway_1990}. Moreover, to uphold the statement that road transportation networks are fragile, an empirical function solely at the link level is far from being sufficient.  

For studying the fragility at the network level, a widely leveraged traffic model is the Macroscopic Fundamental Diagram (MFD), with an illustration in Fig. \ref{fig: demand}. With the assumption of a homogeneous region, an MFD demonstrates the mathematical relationship between the most essential aggregated traffic variables, i.e., flow, density, and speed. Widely applied functional forms of MFDs include polynomials and multi-regime linear functions. Polynomials MFDs are typically fit from field measurements numerically and are commonly seen in research works on traffic control, such as in \cite{kouvelas_enhancing_2017, sirmatel_economic_2018, saeedmanesh_extended_2021, hamedmoghadam_percolation-based_2022}. Alternatively, based on the variation theory \citep{daganzo_variational_2005} and the assumption to simplify a homogenous network into an abstract corridor, \cite{daganzo_analytical_2008} is the first study to generate a multi-regime linear function MFD through an analytical method, which is often referred to as the Method of Cuts (MoC). MoC derives MFDs directly from traffic-related variables with physical meanings, such as free flow speed, backward wave speed, traffic signalization, lane length, average trip length, etc., and thus avoids the complication of installing detectors and massive data gathering. \cite{leclercq_estimating_2013} further improved the original MoC to accommodate topology and signal timing heterogeneity within the network. \cite{saeedmanesh_clustering_2016, ambuhl_approximative_2019, saedi_estimating_2020} also introduced algorithms to partition an entire heterogeneous network into multiple more homogeneous regions. Recent advances on MoC in \cite{tilg_corridor_2023} relax the assumption of demand homogeneity and an abstracted corridor by creating a hypernetwork from a set of corridors and incorporating turn ratios at intersections. Some other analytical methods have also been proposed to produce an MFD, such as through stochastic approximation \citep{laval_stochastic_2015}. However, \cite{tilg_evaluation_2020} has demonstrated that MoC yields a more accurate upper bound for the MFD.

Based on traffic models like MFDs and various performance indicators, different terminologies have been proposed to evaluate the performance of road transportation networks in the past decades, and two commonly used terms to characterize the extent of performance variations under stress are robustness \citep{shang_benchmark_2022} and resilience \citep{mattsson_vulnerability_2015}. Researchers have devoted extensive efforts to the assessment and design of robust and resilient transportation systems of all kinds, such as in railway systems \citep{corman_evaluating_2014}, public transportation \citep{cats_robustness_2016}, and road networks \citep{ampountolas_macroscopic_2017, yang_heterogeneity_2019, leclercq_enforcing_2021}. However, the definitions of robustness and resilience can vary under different contexts, even within the transportation domain itself. \cite{zhou_resilience_2019} provided a synthesis of definitions, wherein robustness involves evaluating a system's ability to maintain its initial state and withstand performance degradation when confronted with uncertainties and disturbances. On the other hand, resilience emphasizes the system's capability and promptness to recover from major disruptions and return to its original state. Nevertheless, both robustness and resilience can overlook the consideration of a longer time span, which is particularly relevant in transportation when accounting for the ever-growing traffic volume in urban road networks and the exponentially escalating adversarial consequences. Thus, it is necessary to introduce a new term to address this gap. 

The concept of antifragility was initially proposed in \cite{taleb_antifragile_2012} and mathematically elaborated in \cite{taleb_mathematical_2013}, which serves as a general framework for risk identification and management. By embracing current risks, a system can potentially leverage and adapt to future risks of greater magnitudes. When employed in the domain of systems and control, antifragility, together with its counterpart, fragility, can be conceptualized as a nonlinear relationship between the performance and the magnitude of disruptions. If the performance is compromised due to unexpected volatile disruptions, the relationship between the loss in performance and the disruptions would be concave for an antifragile system, while being convex or even exponential for a fragile system. Ever since being proposed, antifragility has gained popularity in the risk engineering community \citep{thekdi_integrated_2019, aven_risk_2022} and across multiple disciplines, such as biology \citep{kim_antifragility_2020}, medicine \citep{axenie_antifragile_2022}, energy \citep{coppitters_optimizing_2023}, robotics \citep{axenie_antifragile_2023}, machine learning \citep{pravin_fragility_2024}, and lately in transportation for designing model-free antifragile perimeter control algorithms to tackle potential disruptive events \citep{sun_antifragile_2024}. It should also be highlighted that although systems can be fragile by nature, proper intervention and control strategies can enhance their antifragility against increasing levels of disruptions \citep{axenie_antifragility_2024, axenie_applied_2026}. The concept of antifragility resembles the philosophy of another term that has recently attracted attention in transportation, self-organized criticality \citep{laval_self-organized_2023}, which analyzes traffic flow from a complex systems perspective, implying an avalanche effect when the traffic state ever becomes critical.

This work makes the following contributions by studying the fragile nature of road transportation networks:

\begin{enumerate}
    \item \textit{Proof of concept:} Previous studies on traffic performance showing signs of fragility have primarily relied on empirical data and intuitive reasoning. This research aims to establish the fragile nature of road transportation networks through rigorous mathematical analysis.
    \item \textit{Methodological contribution:} A skewness-based fragility indicator inspired by the sigmoid curve is developed for the approximation of the fragility of a network. A scalable unit MFD can be applied for the cross-comparison of their fragile properties among different networks, relying merely on parameters with physical meanings.
    \item \textit{From theory to reality:} As stochasticity prevails in road transportation systems in the real world, we also designed a numerical simulation calibrated with real-world parameters to study the impact of such uncertainties on the fragile properties of transportation networks.
\end{enumerate}

The overarching objective of this work is to provide insights to transportation researchers for the future design of transportation networks and control strategies to be not only robust and resilient but also antifragile. The remainder of this paper is structured as follows. Section \ref{sec: formulation} formulates the mathematical definition of (anti-)fragility and its detection methods. Then we conduct the mathematical proof to establish the fragile nature in Section \ref{sec: proof}, whereas a skewness-based fragility indicator is proposed in Section \ref{sec: indicator}. Section \ref{sec: numerical} presents the numerical simulation with a real-world network and stochasticity. With Section \ref{sec: conclusion}, we conclude the analysis on the fragile nature of road transportation networks and its implications for future studies.

\section{Problem formulation}
\label{sec: formulation}

\begin{figure}[h!]
\includegraphics[width=0.98\textwidth]{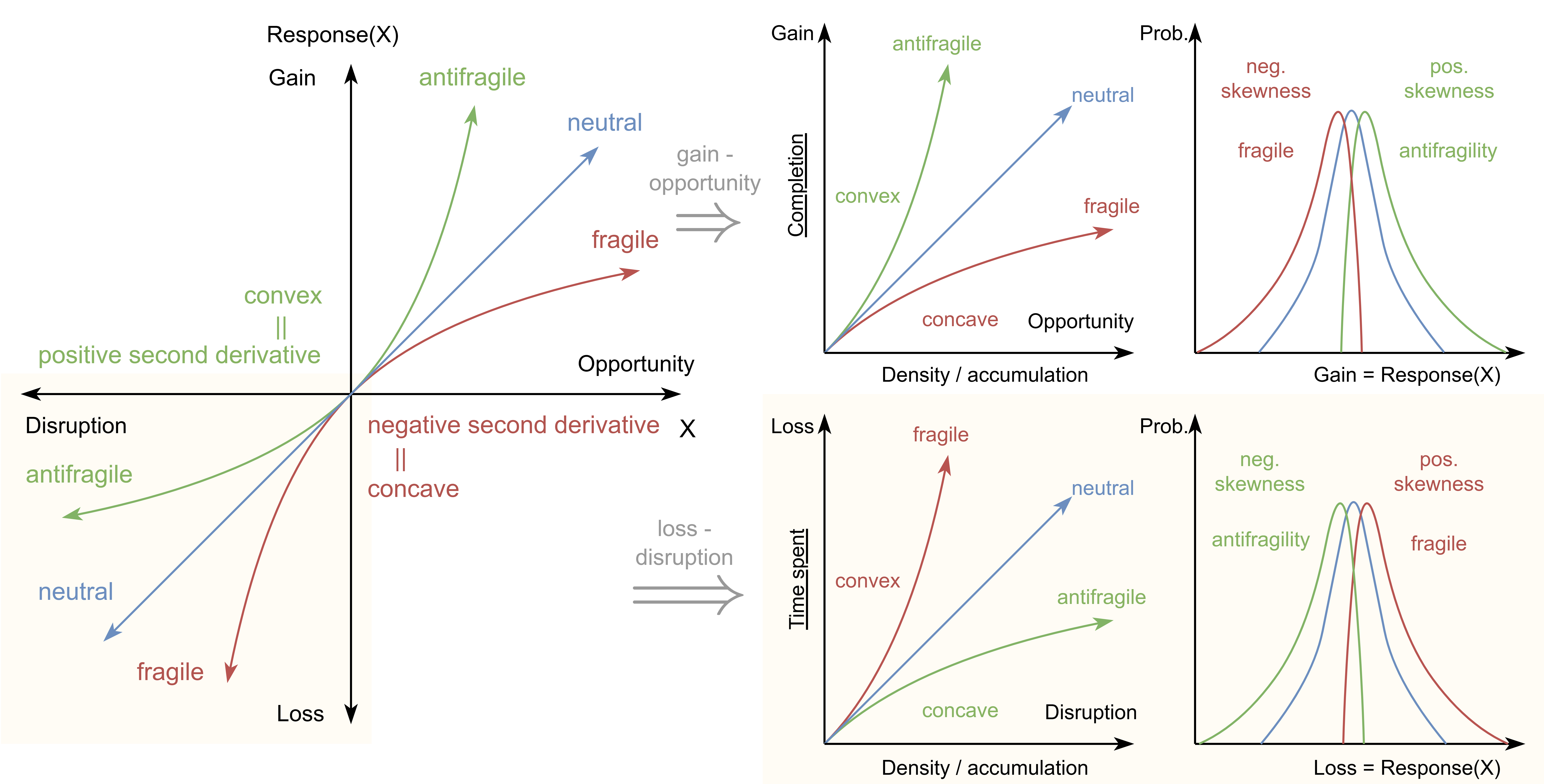}
\centering
\caption{Antifragility in a four-quadrant diagram (The probability density curves serve as an illustration and do not precisely reflect the performance distribution).}
\label{fig: antifragility_all}
\end{figure}

To determine a system's (anti-)fragile properties, the relationship between the system's performance and disruptions has to be identified. In this study, we examine the relationship between vehicle time spent versus disruption magnitudes to validate the fragile nature of road transportation networks. A bigger picture of (anti-)fragility, as shown in Fig. \ref{fig: antifragility_all}, can facilitate understanding of the gist behind it. In this four-quadrant diagram, while the first quadrant depicts a gain-opportunity relationship, the third quadrant demonstrates the loss-disruption relationship, which is the focus of the study. Although convexity signifies an antifragile response on the whole four-quadrant diagram, when it comes to each quadrant, convexity or concavity itself does not necessarily indicate (anti-)fragility without the context. Also, a variable can be an opportunity for one performance indicator but becomes a disruption for another. For example, while traffic density always imposes a non-positive effect on travel speed and thus follows the loss-disruption relationship with the fragile response being convex, it can actually be beneficial to network output so that the related fragility curve is concave. In our focus, as highlighted in beige in Fig. \ref{fig: antifragility_all}, both nonlinear functions can be represented by Jensen's inequality, with either $\mathbb{E}[g(X)] \geq g(\mathbb{E}[X])$ for a fragile response or $\mathbb{E}[g(X)] \leq g(\mathbb{E}[X])$ for an antifragile response. This relationship can then be determined through the second derivative \citep{ruel_jensens_1999}, i.e., a positive second derivative featuring a convex function and hence a fragile system, and vice versa. Note that the calculation of the derivatives is only possible when the function is continuous and differentiable, which means the underlying mathematical model representing the system dynamics needs to be known beforehand. 

As (anti-)fragility represents the asymmetry of a probability distribution between performance response and disruption, skewness as a statistical measure can be applied to further quantitatively evaluate the degree of a system's fragility, as long as the underlying performance function follows convex transformation \citep{zwet1964convex}, with application examples in \cite{taleb_mathematical_2013, coppitters_optimizing_2023}. For a loss-disruption relationship, a positive skewness represented by the long tail pointing to the right indicates a fragile response, whereas negative skewness showcases the fragility of a gain-opportunity relationship. It should be noted, however, that skewness does not strictly indicate (anti-)fragility since certain systems may exhibit a performance function that yields a positive skewness overall but is partially concave and partially convex with disruptions on different domain segments. Therefore, a mathematical analysis of the completion domain is essential if the system's fragility is to be determined.

\begin{figure}[hbt]
  \centering
    \subfigure[Demand disruption]{%
      \resizebox*{8.cm}{!}{\includegraphics{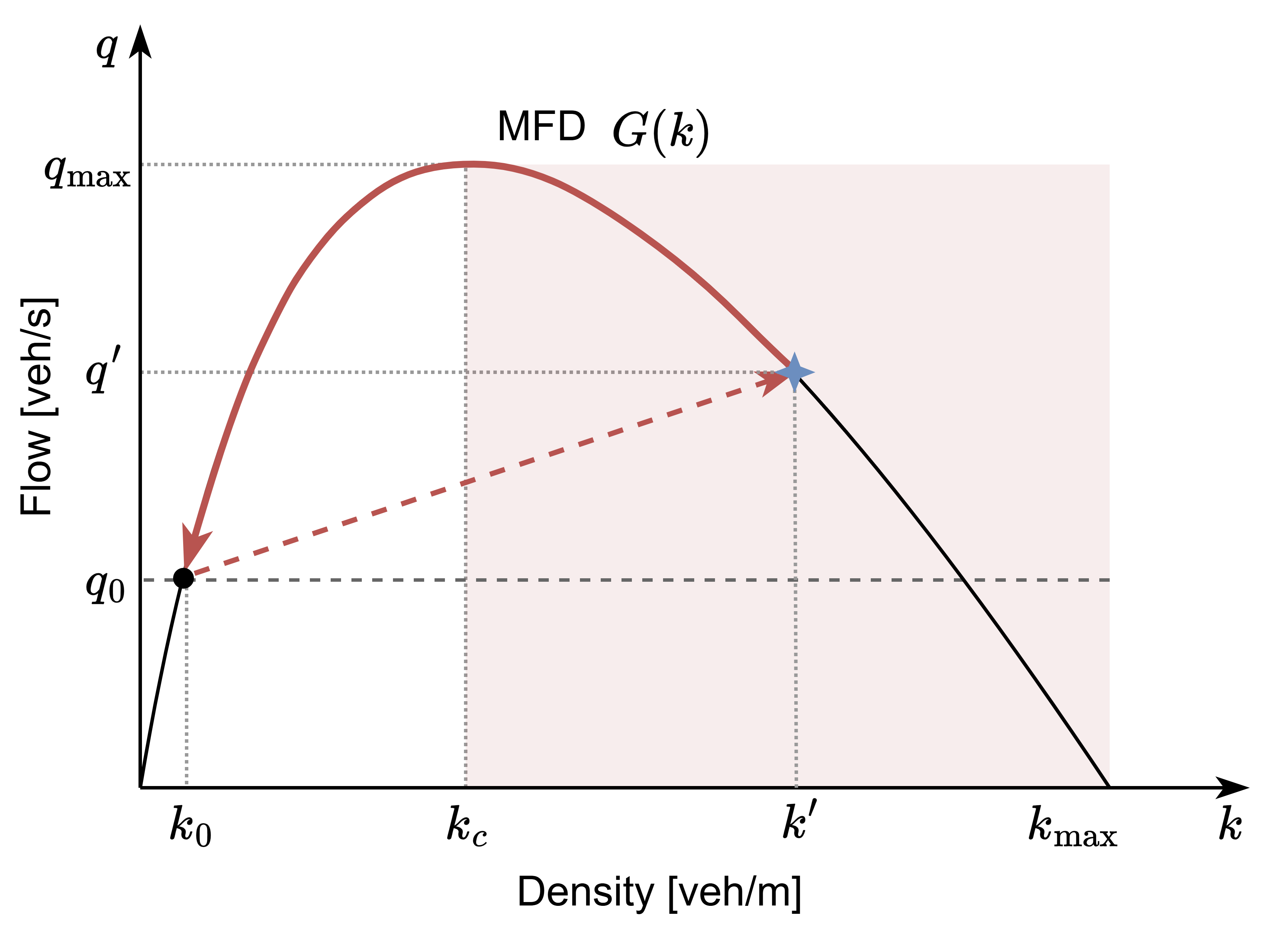}}
      \label{fig: demand}}
      \hspace{5pt}
    \subfigure[Supply disruption]{%
      \resizebox*{8.cm}{!}{\includegraphics{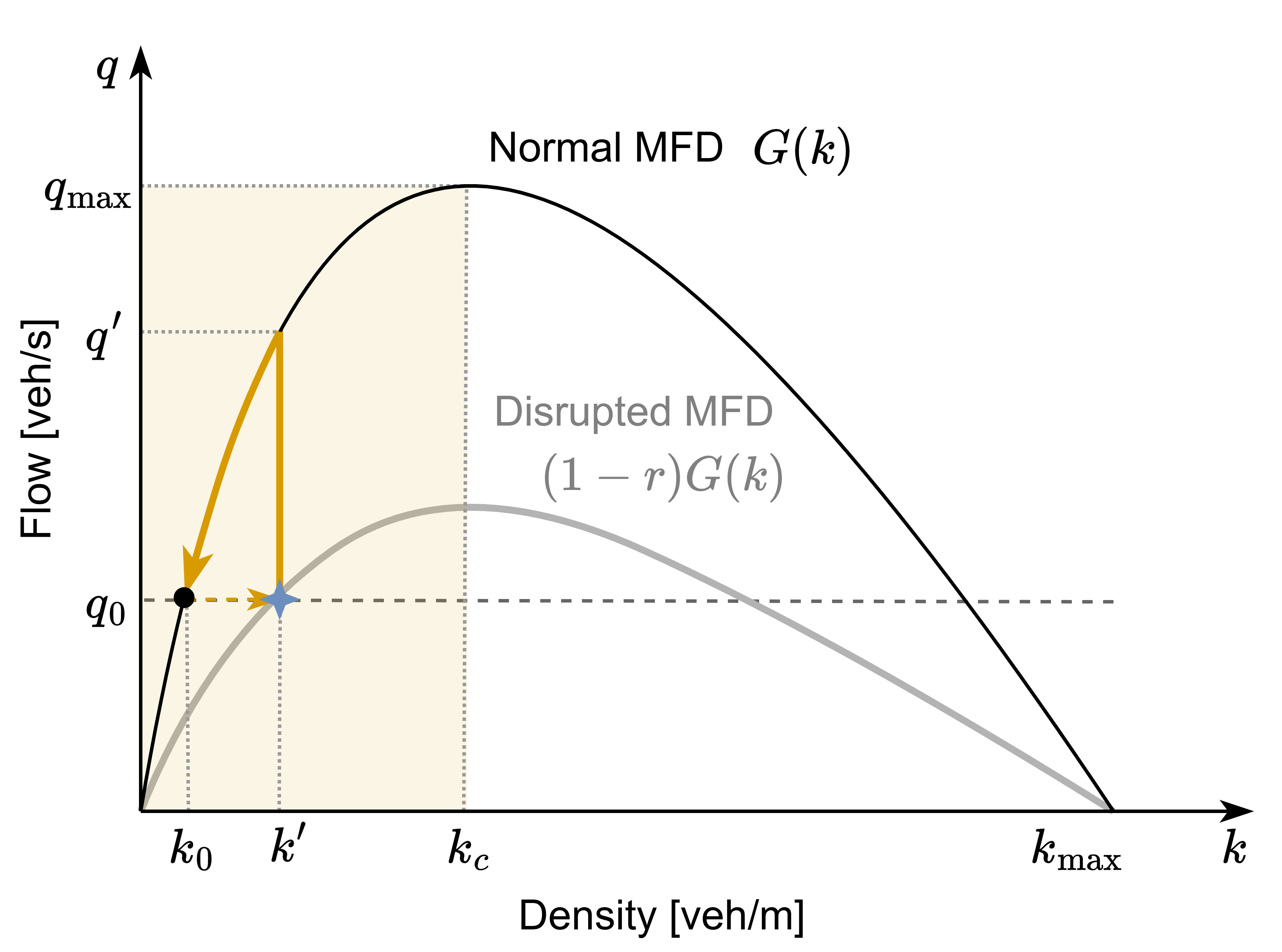}}
      \label{fig: supply}}
    \caption{Disruptions shown on a generic MFD (Black dot: the initial equilibrium without disruption; Dashed line: onset of disruption; Blue 4-point star: new traffic state after disruption; Solid curve: recovery from disruption).} 
    \label{fig: disruption}
\end{figure}

For the following analysis of the fragile nature of road transportation networks, considering that traffic networks primarily involve the management of supply and demand, we distinguish potential disruptions as demand and supply disruptions so that any real-world traffic disruption can be classified as either one. A demand disruption can be easily understood as, for example, surging traffic due to a social event, whereas a supply disruption may indicate an impaired network due to external factors, such as adverse weather or lane closure. As illustrated in Fig. \ref{fig: demand} and Fig. \ref{fig: supply}, we denote a generic MFD profile as $q=G(k)$ representing the relationship between flow $q$ and density $k$ and assume a constant base demand in the network as $q_0$, resulting in an equilibrium traffic state $(k_0, q_0)$ without any disruption. The initial density at equilibrium, the critical density, the new density after disruption, and the gridlock density are denoted as $k_0$, $k_c$, $k'$, and $k_{\rm max}$, respectively. In this case, for the study of demand disruptions, it can be considered that the surging traffic disruption is instantaneous and pushes the traffic state directly to the disruption density $k'$, followed by a gradual recovery process to the normal density. For supply disruptions, we introduce a disruption magnitude coefficient, denoted as $r$, and the disrupted MFD profile can be represented by $(1-r)G(k)$, with the critical density $k_c$ and the jam density $k_{\rm max}$ remaining unchanged. When the supply constraint is lifted, the traffic state will gradually recover to the initial state. Several assumptions need to be further established for the analysis. 

\begin{assumption} 
\label{ass: onset}
This study focuses on the recovery process following disruptions.
\end{assumption}
Since disruptions are abnormal events with different manifestations of onset and can last for an uncertain duration, our study assumes disruptions are instantaneous for simplification. More precisely, a demand disruption instantaneously pushes the original equilibrium to the new disrupted traffic state in Fig. \ref{fig: demand}. Similarly, the removal of supply constraints immediately shifts the disrupted MFD to the normal MFD in Fig. \ref{fig: supply}. We then focus on examining the fragile properties of road transportation networks, targeting the recovery process from either demand or supply disruptions, with the system dynamics explicitly defined by MFDs. 

\begin{assumption} 
\label{ass: below}
The base demand should be sufficiently low. 
\end{assumption}
Following Assumption \ref{ass: onset}, a critical condition to be avoided in this study is the network succumbing to a complete gridlock, where recovery is not possible anymore. If the base demand flow $q_0$ is higher than the demand disruption flow $q'=G(k')$ after a demand disruption or $q'=(1-r)G(k_c)$ after a supply disruption, then the traffic state will gradually move to the gridlock point $(k_{\rm max},0)$. Mathematically, this assumption can be formulated as $q_0 < G(k')$ for demand disruptions as well as $q_0 < (1-r)G(k_c)$ for supply disruptions. 

\begin{assumption} 
\label{ass: focus}
Demand and supply disruptions fall on the opposite sides of the critical density.
\end{assumption}

A surging demand should be considered a disruption only when it pushes the traffic state to the congested branch on the MFD, illustrated as the red shaded area in Fig. \ref{fig: demand}, causing a reduction in the network's maximal possible serviceability. For supply disruption, similar to Assumption \ref{ass: below}, if the traffic state can ever surpass the maximal capacity of the disrupted MFD profile, the traffic density will continue to accumulate until a full gridlock. Therefore, the traffic state after disruption shall still lie within the uncongested zone, shown as the beige shaded area in Fig. \ref{fig: supply}. Mathematically, this assumption can be formulated as $k' > k_c$ for demand disruptions, whereas $k' < k_c$ for supply disruptions. 

Note in Fig. \ref{fig: disruption} we've been using a generic $q-k$ MFD as a representation since it is the more commonly applied form and makes the above assumptions more comprehensible. However, when the total lane length and average trip length, denoted by $D$ and $L$ respectively, are known for a given network, the trip completion, or sometimes referred to as the outflow rate, denoted by $m$, and vehicle accumulation within the network denoted by $n$, can be further determined with Eq. \ref{eq: n} and Eq. \ref{eq: m} \citep{leclercq_estimating_2013, geroliminis_optimal_2013}. While keeping the shape of the MFD profile identical, the advantage of using an $m-n$ MFD is that it can serve as the system dynamics to deduce future system states, which is a prevalent practice in the research of various traffic control strategies, as in \citep{zhou_model-free_2021, kouvelas_linear-parameter-varying_2023}, and is the foundation for the following study.

\begin{subequations}
\begin{align}
\label{eq: n}
n &= kD
\\
\label{eq: m}
m &= \frac{qD}{L}
\end{align}
\end{subequations}

For clarity, a notation list is summarized in Table \ref{tab: notation}, with the mapping of variables between the $q-k$ MFD and the $m-n$ MFD. Another separate table of notations with real-world values for the numerical simulation with stochasticity can be found in Section \ref{sec: numerical}.

\begin{table}[hbt!]
\small
\caption{List of notations.}
\label{tab: notation}
\renewcommand{\arraystretch}{1.2}
\begin{tabular}{L{1.5cm}L{6.5cm}L{1.2cm}L{6.4cm}}
\hline
\multicolumn{4}{l}{\textbf{Common notations}}    
\\\hline
$t$ & \multicolumn{3}{l}{Time}
\\
$TTS$ & \multicolumn{3}{l}{Total time spent}
\\
$L$ & \multicolumn{3}{l}{Average trip length}
\\
$D$ & \multicolumn{3}{l}{Total lane length of a network}
\\
$r$ & \multicolumn{3}{l}{Supply disruption magnitude coefficient}
\\\hline
\multicolumn{4}{l}{\textbf{MFD-related notations}}    
\\\hline
$N$ & \multicolumn{3}{l}{Total number of cuts according to MoC}
\\
$y$ & \multicolumn{3}{l}{Starting cut (the \textbf{most} congested cuts)}
\\
$z$ & \multicolumn{3}{l}{Ending cut (the \textbf{least} congested cuts)}
\\
$i$ & \multicolumn{3}{l}{Sequential number of cuts from \textbf{the most to the least} congested cuts}
\\
\multicolumn{2}{l}{\underline{\underline{$q-k$ MFD}} (flow MFD)} & \multicolumn{2}{l}{\underline{\underline{$m-n$ MFD}} (outflow MFD)}
\\
$G(\cdot)$ & Function of MFD & $M(\cdot)$ & Function of MFD
\\
$k$ & Density & $n$ & Vehicle accumulation 
\\
$k_{c}$ & Critical density & $n_{c}$ & Critical vehicle accumulation
\\
$k_{\rm max}$ & Maximal density at gridlock  & $n_{\rm max}$ & Maximal accumulation at gridlock
\\
$k_0$ & Initial density at equillibrium & $n_0$ & Initial accumulation at equillibrium
\\
$k'$ & Disruption density & $n'$ & Disruption vehicle accumulation
\\
$q$ & Traffic flow & $m$ & Trip completion
\\
$q_{\rm max}$ & Capacity of the network with $k=k_c$ & $m_{\rm max}$ & Maximal trip completion with $n=n_c$
\\
$q_{0}$ & Base demand flow with $k=k_0$ & $m_{0}$ & Base demand inflow with $n=n_0$
\\
$q'$ & Flow at disruption density with $k=k'$ & $m'$ & Completion at disruption with $n=n'$ 
\\
$v$ & Speed with $v=q/k$ & $a$ & Completion rate, $a=m/n$
\\
$u_i$ & Gradient of cut $i$ & $a_i$ & Gradient of cut $i$
\\
$c_i$ & Intercept of cut $i$ & $b_i$ & Intercept of cut $i$
\\
$u_f$ & Free flow speed, $u_f = u_N$ & $a_f$ & Gradient of the free flow cut
\\
$w$ & Backward wave speed, $w = u_1$ & $a_w$ & Gradient of the backward wave cut
\\
$n_{\tilde{c}, i}, m_{\tilde{c}, i}$ & \multicolumn{3}{l}{Critical accumulation and completion between cut $i$ and cut $i+1$ following MoC}
\\
$n_{\check{c}}, m_{\check{c}}$ & \multicolumn{3}{l}{Critical accumulation and completion of the virtual interception between cut $y$ and cut $z$}
\\
$t_{\tilde{c}, i}$ & \multicolumn{3}{l}{Time to reach critical density/accumulation between cut $i$ and cut $i+1$}
\\\hline
\end{tabular}
\end{table}

\section{Mathematical analysis of the fragility of road transportation networks}
\label{sec: proof}

In this section, we conduct a mathematical analysis to evaluate the fragile nature of road transportation networks under either a demand or supply disruption. Total Time Spent (TTS) is applied as the indicator for the examination of system performance, which was first introduced in \cite{edie1965discussion} as the aggregate time spent by all vehicles in passing through a space-time domain. As outlined in Section \ref{sec: formulation}, the presence of a positive second derivative in performance loss regarding the magnitude of disruption indicates the system's fragility. Therefore, to illustrate the transportation network's fragile nature, we analyze the derivatives of TTS over the disruption magnitudes. Suppose the system is neither fragile nor antifragile, then this approach is expected to yield a linearly growing performance loss alongside an increasing magnitude of disruptions, with the second derivative being zero. 

\begin{prop}
\label{prop: 1}
Road transportation networks are fragile under demand disruptions.
\end{prop}

\begin{proof}
To prove that road transportation networks are fragile under demand disruptions, we compute the TTS on each involved cut following a disruption recovery process. Then the derivatives of TTS over the disruption demand $n'$ are calculated and summed, and the demand-side fragility of road transportation networks would be proven if the second derivative $\frac{d^2TTS}{dn'^2}$ were to be positive.

As introduced in Section \ref{sec: introduction}, analytical forms of MFDs have the advantage of being derived from parameters with physical meanings that are related to traffic networks, whereas numerical MFDs have to be approximated from sensor installation and on-site measurements with algebraic functions. Therefore, we use the analytically generated MoC as the foundational model for our subsequent analysis. MoC is composed of a series of linear functions, which are referred to as cuts $1, 2, \dots, N$, with increasing gradients as the vehicle accumulation decreases and forming a concave envelope as presented in Fig. \ref{fig: moc_simplified_re}. For a given cut $i$, the slope and the intercept on the y-axis on the coordinates are denoted as $a_i$ and $b_i$, with $a_{i+1}>a_i$ and $b_i>b_{i+1} \geq 0$. The critical accumulation $n_{\tilde{c}, i}$ on the $x$-axis does not represent the critical point of the entire MFD, but rather the critical accumulation between any two consecutive cuts $i$ and $i+1$ with two exceptional cases being $n_{\tilde{c}, 0}=n_{\rm max}$ and $n_{\tilde{c}, N}=0$. According to Assumption \ref{ass: onset}, we simplify this surging demand as a disruption that takes place instantly in the network, denoted as $n'$ at time $t_0=0$ and lands on cut $y$. At time $t_{\tilde{c}, i} \; (y \leq i < z)$, the number of vehicles in the network reaches the corresponding critical accumulation point $n_{\tilde{c}, i}$ intersected by cut $i$ and $i+1$. And after any period $t$, the traffic state lands on cut $z$. A virtual intersection point $(n_{\check{c}}, m_{\check{c}})$ between cut $y$ and cut $z$ is also illustrated and will be discussed later in this section. We also denote the initial trip completion as $m'=a_y n'+b_y$ and the critical trip completion as $m_{\tilde{c}, i} = a_i n_{\tilde{c}, i}+b_i = a_{i+1} n_{\tilde{c}, i}+b_{i+1}$ respectively. $m'_0$ represents the supremum of $m_0$ and gets arbitrarily close to $m_0$ with sufficient long $t$.

\begin{figure}[hbt!]
\includegraphics[width=0.70\textwidth]{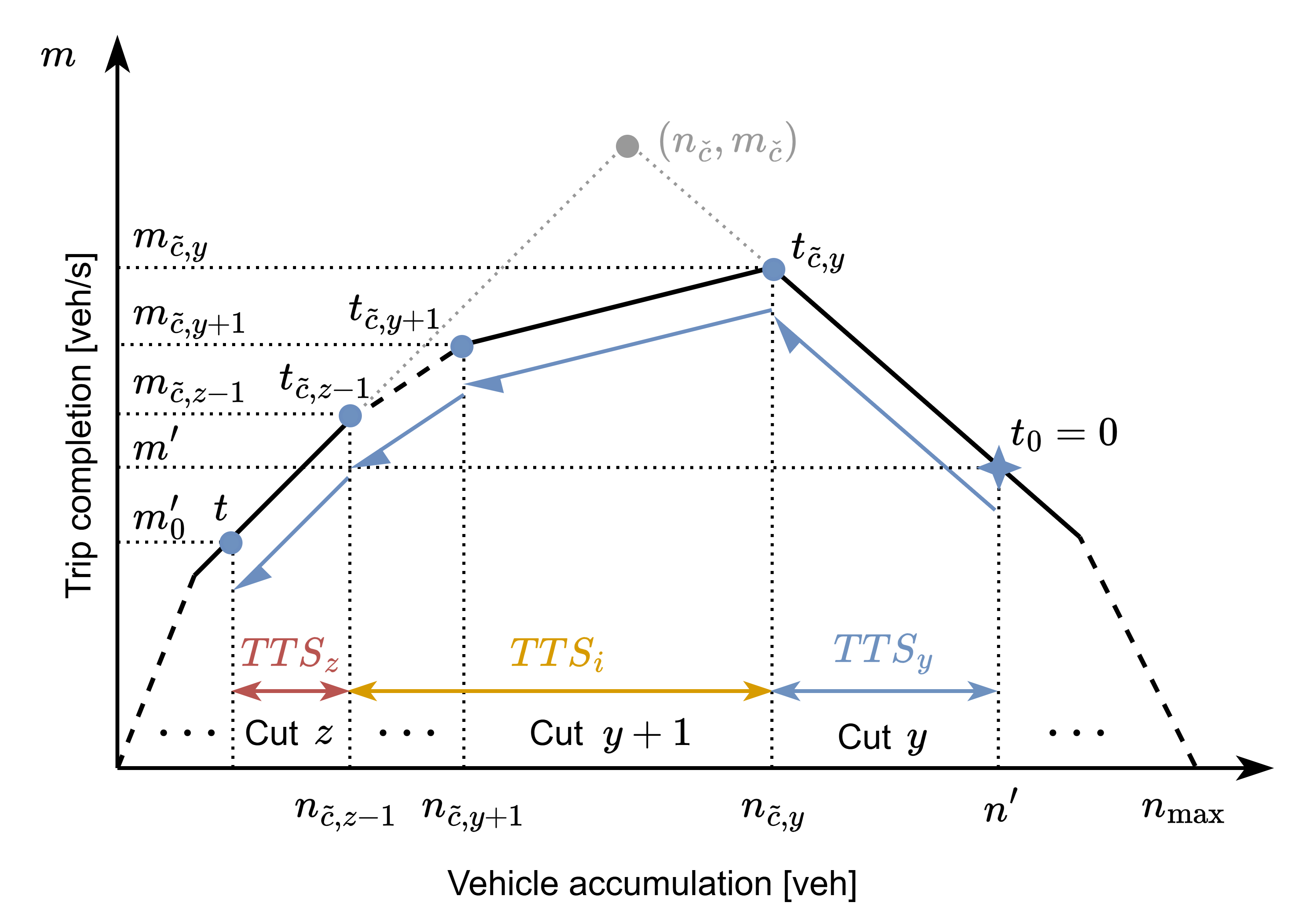}
\centering
\caption{Simplification of MoC under a demand disruption.}
\label{fig: moc_simplified_re}
\end{figure}

Any cut $i$ of the MFD can be formulated as the following Eq. \ref{eq: mfd}:

\begin{equation}
    \label{eq: mfd}
    M(n) = a_i n + b_i, \forall n \in (n_{\tilde{c}, i}, n_{\tilde{c}, i-1}]
\end{equation}

By calculating the difference between the base demand $m_0$ as the inflow rate and the trip completion $M(n)$ as the outflow rate, the system dynamics on cut $i$ can be formulated as Eq. \ref{eq: dynamics}:

\begin{equation}
\label{eq: dynamics}
\frac{dn}{d\tau} = -M(n) + m_0 = - a_in - b_i + m_0
\end{equation}

Now we first assume that the traffic state moves only along a single cut $i$, and with any amount of vehicle accumulation $n_1$ at the beginning of a given period between $t_1$ and $t_2$, the number of vehicles $n_2$ at the end of this period can be determined as:

\begin{subequations}
\begin{alignat}{2}
    && \int_{t_1}^{t_2} d\tau &=- \int_{n_1}^{n_2} \frac{1}{a_in + b_i - m_0} dn
    \\ 
    \label{eq: t_2}
    \Longrightarrow\hspace{16pt} && t_2 - t_1 &= - \frac{1}{a_i} \ln{ \left( \frac{a_in_2+b_i-m_0}{a_in_1+b_i-m_0} \right) }
    \\
    \label{eq: n_2}
    \Longrightarrow\hspace{16pt} && n_2 &= \frac{e^{-a_i(t_2-t_1)}(a_in_1+b_i-m_0)}{a_i}-\frac{b_i-m_0}{a_i}
\end{alignat}
\end{subequations}

Therefore, with the disruption accumulation $n'$, and when the traffic state is assumed to be on the same cut $i$. After any time $t$, the vehicle accumulation $n$ would be:

\begin{align}
n &= \frac{a_i n'+b_i -m_0}{a_i}e^{-a_i t}-\frac{b_i -m_0}{a_i}
\end{align}

The TTS on this single cut $i$ for a given period $t$ can be calculated as:

\begin{align}
    \nonumber
    TTS = \int_{0}^{t} n d\tau &= \int_{0}^{t} \left( \frac{a_i n'+b_i -m_0}{a_i } e^{-a_i \tau} - \frac{b_i -m_0}{a_i } \right) d\tau
    \\
    \label{eq: TTS}
    &= - \frac{a_i n'+b_i -m_0}{a_i ^2} e^{-a_i t} - \frac{b_i -m_0}{a_i }t + \frac{a_i n'+b_i -m_0}{a_i ^2}
\end{align}

Now we compute the derivatives of TTS on the same cut $i$, i.e., $t<t_{\tilde{c}, i}$.

\begin{subequations}
\begin{align}
    \label{eq: first_derivative}
    \frac{dTTS}{dn'} &=  \frac{1}{a_i} - \frac{e^{-a_i t}}{a_i}
    \\
    \frac{d^2TTS}{dn'^2} &=  0
\end{align}
\end{subequations}

The second derivative of TTS is $0$, indicating that when the traffic states move only along a single cut $i$, it shows neither fragility nor antifragility. 

When the traffic state goes over the critical vehicle accumulation $n_{\tilde{c}, i}$, and since the MoC is a piecewise function and thus an additive process, we can calculate the TTS and the related derivatives separately on each concerned cut $y, y+1, \dots, z$, denoted as $TTS_y, TTS_{y+1}, \dots, TTS_z$. The individual TTS are grouped as $TTS_y, TTS_{i}, TTS_z$ with three colors in Fig. \ref{fig: moc_simplified_re}, which will be introduced shortly afterward. Since the time to reach the $i$-th critical point $t_{\tilde{c}, i}$ is yet unknown, we first need to determine $t_{\tilde{c}, i}$ for each critical point based on Eq. \ref{eq: t_2}.

\begin{subequations}
\begin{align}
    \label{eq: t_c}
    t_{\tilde{c}, y} &=  - \frac{1}{a_y} \ln{ \left( \frac{a_y n_{\tilde{c}, y}+b_y-m_0}{a_y n'+b_y-m_0} \right)}
    \\
    \label{eq: t_z_between}
    t_{\tilde{c}, y+1} - t_{\tilde{c}, y} &=  - \frac{1}{a_{y+1}} \ln{ \left( \frac{a_{y+1} n_{\tilde{c}, {y+1}}+b_{y+1}-m_0}{a_{y+1} n_{\tilde{c}, y}+b_{y+1}-m_0} \right)}
    \\
    &\;\;\vdots \notag
    \\
    \label{eq: t_z_1}
    t_{\tilde{c}, z-1} - t_{\tilde{c}, z-2} &=  - \frac{1}{a_{z-1}} \ln{ \left( \frac{a_{z-1} n_{\tilde{c}, {z-1}}+b_{z-1}-m_0}{a_{z-1} n_{\tilde{c}, z-2}+b_{z-1}-m_0} \right)}
\end{align}
\end{subequations}

Therefore, the time $t_{\tilde{c}, z-1}$ to reach the last critical point $n_{\tilde{c}, z-1}$ can be obtained by summing Eq. \ref{eq: t_c}, \ref{eq: t_z_between}, and \ref{eq: t_z_1} all together:

\begin{align}
    \label{eq: t_z}
    t_{\tilde{c}, z-1} &=  - \frac{1}{a_y} \ln{ \left( \frac{a_y n_{\tilde{c}, y}+b_y-m_0}{a_y n'+b_y-m_0} \right)} - \sum_{i=y+1}^{z-1} \frac{1}{a_{i}} \ln{ \left( \frac{a_{i} n_{\tilde{c}, {i}}+b_{i}-m_0}{a_{i} n_{\tilde{c}, i-1}+b_{i}-m_0} \right)}
\end{align}

As $a_y n_{\tilde{c}, y}+b_y$ is equal to $m_{\tilde{c}, y}$, we can rewrite the above Eq. \ref{eq: t_c}, Eq. \ref{eq: t_z_between} to \ref{eq: t_z_1} as a generalized form, and Eq. \ref{eq: t_z} each as:

\begin{subequations}
\begin{align}
    \label{eq: t_c_m}
    t_{\tilde{c}, y} =& - \frac{1}{a_y} \ln{ \left( \frac{m_{\tilde{c}, y}-m_0}{a_y n'+b_y-m_0} \right)}
    \\
    \label{eq: t_c_t_c_1}
    t_{\tilde{c}, i} - t_{\tilde{c}, i-1} =& - \frac{1}{a_{i}} \ln{ \left( \frac{ m_{\tilde{c}, {i}}-m_0}{m_{\tilde{c}, i-1}-m_0} \right)}
    \\
    \label{eq: t_z_m}
    t_{\tilde{c}, z-1} =& - \frac{1}{a_y} \ln{ \left( \frac{m_{\tilde{c}, y}-m_0}{a_y n'+b_y-m_0} \right)} - \sum_{i=y+1}^{z-1} \frac{1}{a_{i}} \ln{ \left( \frac{m_{\tilde{c}, i}-m_0}{m_{\tilde{c}, i-1}-m_0} \right)}
\end{align}
\end{subequations}

Now we calculate the TTS in the first group $TTS_y$, which contains only the most congested cut $y$ in Fig. \ref{fig: moc_simplified_re}. As cut $y$ ends at $t_{\tilde{c}, y}$, we substitute $t$ in Eq. \ref{eq: TTS} with $t_{\tilde{c}, y}$ in Eq. \ref{eq: t_c_m}, and the $TTS_y$ for cut $y$ will be:

\begin{align}
    \nonumber
    TTS_y=  \int_{0}^{t_{\tilde{c}, y}} n d\tau  &= - \frac{a_y n'+b_y -m_0}{a_y^2} e^{-a_y t_{\tilde{c}, y}} - \frac{b_y-m_0}{a_y}t_{\tilde{c}, y} + \frac{a_y n'+b_y-m_0}{a_y^2}
    \\
    &= - \frac{m_{\tilde{c}, y}-m_0}{a_y^2} + \frac{b_y-m_0}{a_y^2} \ln{ \left( \frac{m_{\tilde{c}, y}-m_0}{a_yn'+b_y-m_0} \right) } + \frac{a_yn'+b_y-m_0}{a_y^2}
\end{align}

The derivatives for $TTS_y$ over the disruption accumulation $n'$ are:

\begin{subequations}
\begin{align}
    \label{eq: dTTS_y}
    \frac{dTTS_y}{dn'} &=  -\frac{b_y-m_0}{a_y}(a_yn'+b_y-m_0)^{-1} + \frac{1}{a_y}
    \\
    \label{eq: d2TTS_y}
    \frac{d^2TTS_y}{dn'^2} &=  (b_y-m_0)(a_yn'+b_y-m_0)^{-2}
\end{align}
\end{subequations}

Then we consider the second group of cuts embraced by the yellow arrow in Fig. \ref{fig: moc_simplified_re}, which comprises all the cuts in between the most congested cut $y$ and the least congested cut $z$. The traffic state on these intermediate cuts covers the full range from one end to the other, and are significantly distinct from the first and the last cuts, where the traffic state can begin or end midway. Combining Eq. \ref{eq: n_2} and Eq. \ref{eq: TTS}, the $TTS_i$ on any intermediate cut $y<i<z$ is:

\begin{align}
    \nonumber
    TTS_i=  \int_{t_{\tilde{c}, i-1}}^{t_{\tilde{c}, i}} n d\tau  &= \int_{t_{\tilde{c}, i-1}}^{t_{\tilde{c}, i}} \left( \frac{e^{-a_i (\tau-t_{\tilde{c}, i-1})} (a_i n_{\tilde{c}, i-1}+b_i -m_0)}{a_i}  - \frac{b_i -m_0}{a_i } \right) d\tau
    \\
    \label{eq: TTS_i}
    &= -\frac{e^{-a_i (t_{\tilde{c}, i}-t_{\tilde{c}, i-1})} (m_{\tilde{c}, i-1} -m_0)}{a_i^2} + \frac{(m_{\tilde{c}, i-1} -m_0)}{a_i^2} - \frac{b_i -m_0}{a_i }(t_{\tilde{c}, i}-t_{\tilde{c}, i-1})
\end{align}

Then $t_{\tilde{c}, i}-t_{\tilde{c}, i-1}$ in the above Eq. \ref{eq: TTS_i} can be substituted with Eq. \ref{eq: t_c_t_c_1}, and we get:

\begin{align}
    \label{eq: TTS_i_new}
    TTS_i= -\frac{m_{\tilde{c}, i}-m_{\tilde{c}, i-1}}{a_i^2} + \frac{b_i -m_0}{a_i^2} \ln{ \left( \frac{ m_{\tilde{c}, {i}}-m_0}{m_{\tilde{c}, i-1}-m_0} \right)}
\end{align}

Since Eq. \ref{eq: TTS_i_new} is independent of $n'$. Therefore, both the first and second derivative of $TTS_i$ over $n'$ with $y<i<z$ are:

\begin{subequations}
\begin{align}
    \label{eq: dTTS_i}
    \frac{dTTS_i}{dn'} &=  0
    \\
    \label{eq: d2TTS_i}
    \frac{d^2TTS_i}{dn'^2} &=  0
\end{align}
\end{subequations}

The last component to consider is the least congested cut $z$, represented by the red arrow in Fig. \ref{fig: moc_simplified_re}. Similar to Eq. \ref{eq: TTS_i}, the $TTS_z$ is:

\begin{align}
    \label{eq: TTS_z}
    TTS_z= -\frac{e^{-a_z (t-t_{\tilde{c}, z-1})} (m_{\tilde{c}, z-1} -m_0)}{a_z^2} + \frac{(m_{\tilde{c}, z-1} -m_0)}{a_z^2} - \frac{b_z -m_0}{a_z }(t-t_{\tilde{c}, z-1})   
\end{align}

This time we substitute $t_{\tilde{c}, z-1}$ in the above Eq. \ref{eq: TTS_z} with Eq. \ref{eq: t_z_m} and we get:

\begin{align}
    \nonumber
    TTS_z =& -\frac{e^{-a_z t} }{a_z^2} (m_{\tilde{c}, z-1} -m_0) 
    \left( \frac{m_{\tilde{c}, y} - m_0} {a_y n'+b_y-m_0} \right) ^{-\frac{a_z}{a_y}}
    \prod_{i=y+1}^{z-1}\left( \frac{m_{\tilde{c}, i}-m_0} {m_{\tilde{c}, i-1}-m_0} \right) ^{-\frac{a_z}{a_i}} + \frac{(m_{\tilde{c}, z-1} -m_0)}{a_z^2}
    \\
    & - \frac{b_z -m_0}{a_z}\left(t + \frac{1}{a_y} \ln{ \left( \frac{a_y n_{\tilde{c}, y}+b_y-m_0}{a_y n'+b_y-m_0} \right)} + \sum_{i=y+1}^{z-1} \frac{1}{a_{i}} \ln{ \left( \frac{a_{i} n_{\tilde{c}, {i}}+b_{i}-m_0}{a_{i} n_{\tilde{c}, i-1}+b_{i}-m_0} \right)}\right)
\end{align}

Since $m_{\tilde{c}, y} - m_0$, $m_{\tilde{c}, i}-m_0$, and $m_{\tilde{c}, i-1}-m_0$ are all positive constants. For conciseness, we introduce a positive constant $P$:

\begin{align}
    \label{eq: P}
    P = (m_{\tilde{c}, y} - m_0)^{-\frac{a_z}{a_y}}\prod_{i=y+1}^{z-1}\left( \frac{m_{\tilde{c}, i}-m_0} {m_{\tilde{c}, i-1}-m_0} \right) ^{-\frac{a_z}{a_i}} > 0
\end{align}

And now the derivatives for the least congested cut $z$ are:

\begin{subequations}
\begin{align}
    \label{eq: dTTS_z}
    \frac{dTTS_z}{dn'} &= -\frac{e^{-a_z t} P}{a_z} (m_{\tilde{c}, z-1} -m_0)  (a_y n'+b_y-m_0) ^{\frac{a_z}{a_y}-1}
    + \frac{b_z-m_0}{a_z} (a_yn'+b_y-m_0)^{-1}
    \\
    \label{eq: d2TTS_z}
    \frac{d^2TTS_z}{dn'^2} &= -\left(\frac{e^{-a_z t}P}{a_z} (a_z-a_y) (m_{\tilde{c}, z-1} -m_0)  (a_y n'+b_y-m_0) ^{\frac{a_z}{a_y}} + \frac{a_y(b_z-m_0)}{a_z}\right) (a_yn'+b_y-m_0)^{-2}
\end{align}
\end{subequations}

The second derivative of the whole process $\frac{d^2TTS}{dn'^2}$ is the sum of the second derivatives for each of the three groups in Fig. \ref{fig: moc_simplified_re}, i.e., $\frac{d^2TTS_y}{dn'^2}$, ${\sum_{i=y+1}^{z-1}\frac{d^2TTS_i}{dn'^2}}=0$, and $\frac{d^2TTS_z}{dn'^2}$, which have been computed in Eq. \ref{eq: d2TTS_y}, Eq. \ref{eq: d2TTS_i}, and Eq. \ref{eq: d2TTS_z} separately:

\begin{align}
    \nonumber
    \label{eq: sum}
     \frac{d^2TTS}{dn'^2} &= \frac{d^2TTS_y}{dn'^2} + {\sum_{i=y+1}^{z-1}\frac{d^2TTS_i}{dn'^2}} + \frac{d^2TTS_z}{dn'^2}
    \\
    & = \left(b_y-m_0 - \frac{e^{-a_z t}P}{a_z} (a_z-a_y) (m_{\tilde{c}, z-1} -m_0) (m'-m_0) ^{\frac{a_z}{a_y}} - \frac{a_y(b_z-m_0)}{a_z}\right) (m'-m_0)^{-2}
\end{align}

As per Assumption \ref{ass: below}, it is assumed that $m'-m_0>0$ to avoid a complete gridlock in the network. Therefore, if a transportation system is to be fragile, $\frac{d^2TTS}{dn'^2}$ should be positive, and we only need to prove the first term of Eq. \ref{eq: sum} to be positive: 

\begin{align}
    \label{eq: to_prove}
    b_y-m_0 - \frac{e^{-a_z t}P}{a_z} (a_z-a_y) (m_{\tilde{c}, z-1} -m_0) (m'-m_0) ^{\frac{a_z}{a_y}} - \frac{a_y(b_z-m_0)}{a_z} > 0
\end{align}

However, the sign of Eq. \ref{eq: to_prove} cannot be directly determined, so we focus on finding its strict lower bound and analyzing the sign of its infimum. Since $t>t_{\tilde{c}, z-1}$, the following function is strictly increasing in $t$ and the relationship holds regardless of whether $a_z$ is positive or negative: 

\begin{align}
    -\frac{e^{-a_z t}}{a_z} > -\frac{e^{-a_z t_{\tilde{c}, z-1}}}{a_z}
\end{align}

Since all of the following terms $P$, $a_z-a_y$, $m_{\tilde{c}, z-1}-m_0$, and $(m'-m_0) ^{\frac{a_z}{a_y}}$ are all positive, we found the following expression as the lower bound of Eq. \ref{eq: to_prove}, and it is never attained on the domain $t>t_{\tilde{c}, z-1}$:

\begin{align}
    \nonumber
    & b_y-m_0 - \frac{e^{-a_z t}P}{a_z} (a_z-a_y) (m_{\tilde{c}, z-1} -m_0) (m'-m_0) ^{\frac{a_z}{a_y}} - \frac{a_y(b_z-m_0)}{a_z} >
    \\
    \label{eq: inequality}
    & b_y-m_0 - \frac{e^{-a_z t_{\tilde{c}, z-1}}P}{a_z} (a_z-a_y) (m_{\tilde{c}, z-1} -m_0) (m'-m_0) ^{\frac{a_z}{a_y}} - \frac{a_y(b_z-m_0)}{a_z}
\end{align}

We substitute $t_{\tilde{c}, z-1}$ in Eq. \ref{eq: inequality} with Eq. \ref{eq: t_z_m} and the constant $P$ with Eq. \ref{eq: P} and here goes:

\begin{align}
    \label{eq: after_t_c}
    b_y-m_0 - \frac{(a_z-a_y)(m_{\tilde{c}, z-1}-m_0)}{a_z} - \frac{a_y(b_z-m_0)}{a_z} = 
    b_y - m_{\tilde{c}, z-1} - \frac{a_y}{a_z} (b_z - m_{\tilde{c}, z-1})
\end{align}

$m_{\tilde{c}, z-1}$ is the trip completion rate when the vehicle accumulation is $n_{\tilde{c}}$ on cut $z-1$ and can be substituted by $a_z n_{\tilde{c}, z-1} + b_z$, so Eq. \ref{eq: after_t_c} can be further simplified as:

\begin{align}
    \label{eq: after_t_c_simp}
    b_y - a_z n_{\tilde{c}, z-1} - b_z - \frac{a_y}{a_z} (b_z - a_z n_{\tilde{c}, z-1} - b_z) = b_y - b_z + (a_y - a_z)n_{\tilde{c}, z-1}
\end{align}

As cut $z$ is less congested than cut $y$, both $b_y>b_z$ and $a_y<a_z$ hold true. In an extreme case where cut $y$ and cut $z$ are consecutive, i.e., $z = y+1$, the virtual intersection point exists physically $n_{\check{c}}= n_{\tilde{c}, z-1}$. Otherwise, when there is any other cut in between, then $n_{\check{c}}> n_{\tilde{c}, z-1}$ holds. Therefore, it can be formulated as $n_{\check{c}} \geq n_{\tilde{c}, z-1}$ and we define a non-negative value $\Delta n = n_{\check{c}} - n_{\tilde{c}, z-1}$ so that Eq. \ref{eq: after_t_c_simp} can then be rewritten as:

\begin{align}
    \nonumber
    \label{eq: n_c_check}
    b_y - b_z + (a_y - a_z) (n_{\check{c}} - \Delta n) &= (a_yn_{\check{c}}+b_y)-(a_zn_{\check{c}}+b_z) - (a_y - a_z)\Delta n
    \\
    &=m_{\check{c}} - m_{\check{c}} - (a_y - a_z)\Delta n \geq 0
\end{align}

Having demonstrated its infimum to be non-negative, it is certain that:

\begin{align}
     \frac{d^2TTS}{dn'^2} &=  \left( b_y-m_0 - \frac{e^{-a_z t}P}{a_z} (a_z-a_y) (m_{\tilde{c}, z-1} -m_0) (m'-m_0) ^{\frac{a_z}{a_y}} - \frac{a_y(b_z-m_0)}{a_z}\right) (m'-m_0)^{-2} > 0
\end{align}

The second derivative of TTS over the disruption vehicle accumulation $n'$ is positive, which indicates the fragility of road transportation networks under demand disruption.
\end{proof}

While a positive second derivative of TTS over traffic demand signifies the fragility of the road transportation network to demand disruptions, establishing a positive second derivative of TTS concerning the magnitude of MFD disruption would demonstrate the fragility from the perspective of supply disruptions. As introduced in Section \ref{sec: formulation}, a supply disruption magnitude coefficient $r$ is applied so that the disrupted MFD profile can be expressed as $(1-r)M(n)$. Although real-world MFDs may be impaired in various shapes, we use this simple approach as applied in \cite{ambuhl_functional_2020} when studying the uncertainties of MFDs. The physical meaning of $(1-r)M(n)$ relates to the decrease of the free-flow speed and backward wave speed due to, e.g., snowy weather and icy roads, with the maximal density of the network remaining unchanged. 

\begin{prop}
\label{prop: 2}
Road transportation systems are fragile under supply disruptions.
\end{prop}

\begin{proof}

Fig. \ref{fig: supply_recovery} illustrates the supply disruption recovery process with two potential scenarios: (1) a single cut is involved, represented by $M_1(n)$ and the gray dashed lines, and (2) the traffic state spans multiple cuts, represented by $M_2(n)$ and the solid black lines. The traffic demand at equilibrium before the MFD disruption is $m_0 = M(n_0)$. After the onset of a supply disruption, as per Assumption \ref{ass: below}, the supply disruption magnitude coefficient $r\in[0,1)$ is not exceedingly large, so the demand $m_0$ should remain below the maximal capacity on the disrupted MFD profile $(1-r)M(n_c)$. Therefore, only the cuts below the critical accumulation $n_c$ are presented. The new traffic state after disruption is $m_0 = m' = (1-r)M(n')$. When the supply constraint is lifted, the traffic state recovers to its original state $(n_0,m_0)$ following either the light amber lines if only one single cut is involved or the dark amber lines if it travels through multiple cuts. 

\begin{figure}[hbt!]
\includegraphics[width=0.52\textwidth]{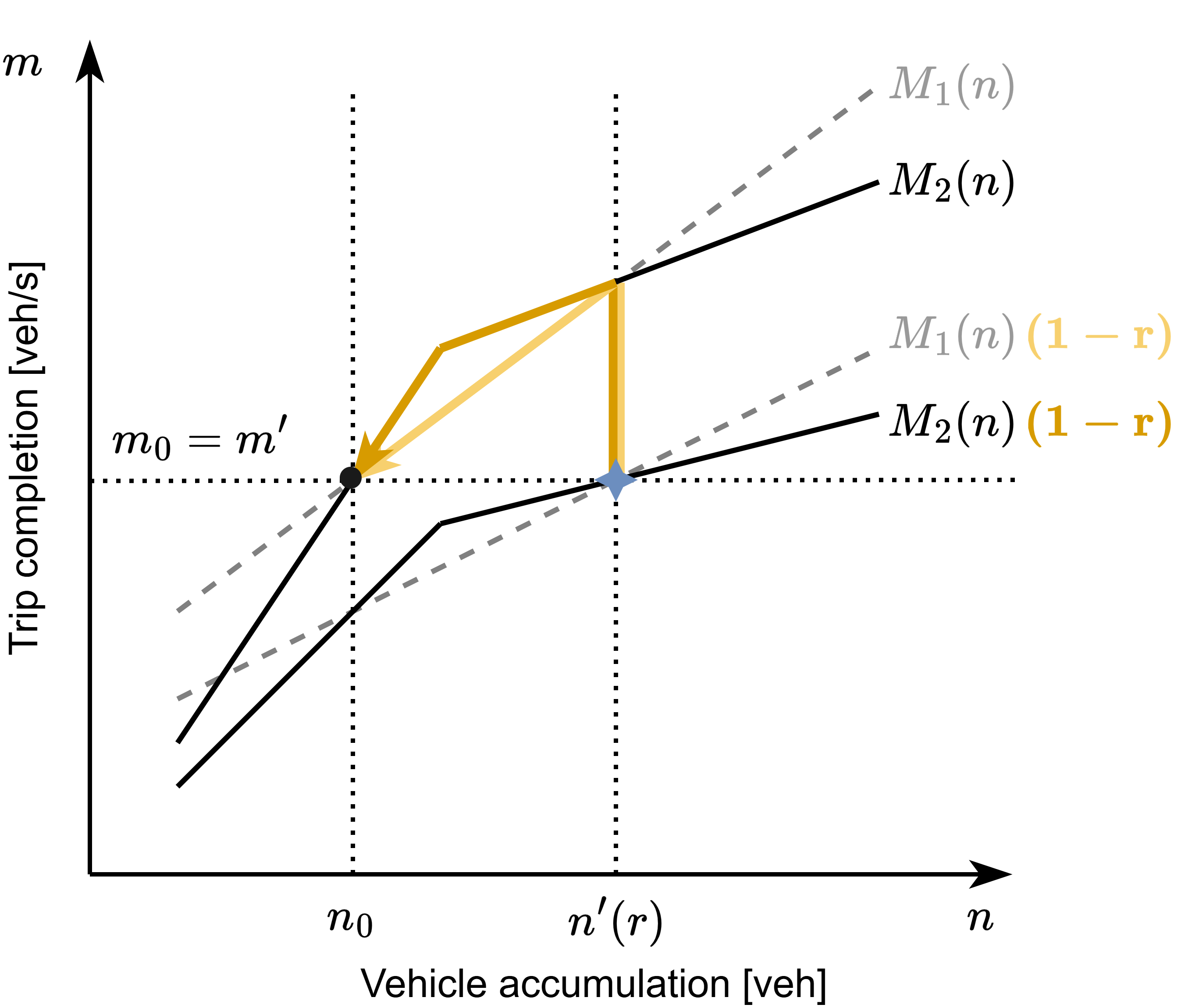}
\centering
\caption{Traffic state recovering from a supply disruption.}
\label{fig: supply_recovery}
\end{figure}

Unlike in the study of demand disruptions, where the disruption demand $n'$ is an independent variable, $n'$ is dependent on $r$ when analyzing supply disruptions. And since TTS is a function of $n'$ and $n'$ is again a function of $r$, by applying the chain rule, we can compute the second derivative of TTS over $r$ as: 

\begin{align}
    \nonumber
    \frac{d^2TTS}{dr^2} &= \frac{d}{dr} \left( \frac{dTTS}{dn'}\cdot \frac{dn'}{dr} \right)
    \\
    \label{eq: supply_recovery}
    &= \frac{d^2TTS}{dn'^2}\cdot \left( \frac{dn'}{dr} \right) ^2 + \frac{dTTS}{dn'}\cdot \frac{d^2n'}{dr^2}
\end{align}

First we demonstrate that both $\frac{dn'}{dr}$ and $\frac{d^2n'}{dr^2}$ are positive when multiple cuts are involved. With the base demand $m_0$ before and after supply disruption being constant, we obtain the following Eq. \ref{eq: supply_m0} considering the traffic state covers multiple cuts, with $y$ and $z$ each denoting the first and the last involved cut with $y \neq z$.

\begin{subequations}
\begin{align}
    \label{eq: supply_m0}
    m_0 &= a_z n_0+b_z = (1-r)(a_y n'+b_y)
    \\
    \Longrightarrow\hspace{16pt} n' &= \frac{a_z n_0+b_z}{a_y (1-r)}-\frac{b_y}{a_y} 
\end{align}
\end{subequations}

The first and second derivatives of $n'$ over the supply disruption magnitude coefficient $r$ can be computed as:

\begin{subequations}
\begin{align}
    \frac{dn'}{dr} &= \frac{a_z n_0+b_z}{a_y}(1-r)^{-2}
    \\
    \frac{d^2n'}{dr^2} &= \frac{2(a_z n_0+b_z)}{a_y}(1-r)^{-3}
\end{align}
\end{subequations}

Since the gradient $a_y$ and trip completion $a_z n_0+b_z$ are both positive, the first and second derivatives of $n'$ over $r$ are positive as well. Additionally, when the traffic state moves along one single cut with $y=z$, the same conclusion still holds: 

\begin{subequations}
\begin{align}
    \frac{dn'}{dr} &= \frac{a_z n_0+b_z}{a_z}(1-r)^{-2}
    \\
    \frac{d^2n'}{dr^2} &= \frac{2(a_z n_0+b_z)}{a_z}(1-r)^{-3}
\end{align}
\end{subequations}

In Proposition \ref{prop: 1}, we've already proven the recovery process to be fragile when the traffic state shifts from a more congested cut to a less congested cut on the MoC MFD, i.e., $\frac{d^2TTS}{dn'^2} > 0$, or to be neither fragile nor antifragile when it stays only on one single cut, i.e., $\frac{d^2TTS}{dn'^2} = 0$. Following Assumption \ref{ass: focus} that $a_z>0$ holds as the cut is below the critical density, through the same procedure as proving the second derivative to be positive, by summing Eq. \ref{eq: dTTS_y}, Eq. \ref{eq: dTTS_i}, and Eq. \ref{eq: dTTS_z}, we can easily prove that the first derivative $\frac{dTTS}{dn'}>0$ when the traffic state travels across multiple cuts in the recovery process following a supply disruption. In addition, even if it stays on one single cut, $\frac{dTTS}{dn'}$ is still positive as demonstrated in Eq. \ref{eq: first_derivative} due to $a_z>0$. 

Since $\frac{d^2TTS}{dn'^2}$ is non-negative, and $\frac{dn'}{dr}$, $\frac{d^2n'}{dr^2}$, as well as $\frac{dTTS}{dn'}$ in Eq. \ref{eq: supply_recovery} have all been demonstrated to be positive, thus $\frac{d^2TTS}{dr^2}$ is always positive, and we've proven the fragile nature of road transportation networks under supply disruptions.
\end{proof}

\section{Fragility indicator and implications}
\label{sec: indicator}

Although road transportation networks have been mathematically proven to be intrinsically fragile, a quantitative approach to assess fragility across various networks has yet to be developed. It can also be of great interest to reveal factors contributing to or mitigating the fragile property of a road network. In Section \ref{sec: formulation}, two generic approaches, namely, the second derivative and the distribution skewness, have been discussed for the evaluation of system fragility. Despite being proven positive and thus showcasing fragile characteristics, the exact values of the second derivative vary even for different traffic states on the same MFD, let alone a cross-comparison among MFDs of various networks. Therefore, we resort to a skewness-based fragility indicator, which allows for the generation of a single fixed value for a region based on its performance density distribution under multiple disruption events of varying magnitudes. For a given network, without considering hysteresis and other sources of uncertainties, a well-defined MFD can be produced and is widely utilized in the research of traffic control \citep{yildirimoglu_approximating_2014}. Assuming no alteration in the driving behavior of the population, the well-defined MFD is primarily considered to be solely dependent on the network topology, including signalization and other potential traffic control strategies \citep{leclercq_estimating_2013}. Therefore, in this section, using only the MFD-related parameters obtained through the analytical traffic model, we propose a skewness-based indicator to evaluate the fragility of a given network. This approach eliminates the need for massive traffic sensor data from sources such as loop detectors, allowing for a pre-assessment of the network fragility during the design phase.

\subsection{Empirical study on the skewness}

Although typically more than three cuts can be generated to form an MFD through MoC, researchers sometimes simplify MFDs to be a trapezoidal shape with one forward, one stationary, and one backward cut \citep{haddad_stability_2012, tilg_evaluation_2020, lee_effects_2023}. In this section, we follow the same simplification to facilitate creating a so-called unit MFD, similar to the isosceles FD proposed in \cite{laval_self-organized_2023}, for the evaluation of fragility, formulated as $m=M(n, a_f, a_w, m_{\rm max}, n_{\rm max})$, where $a_f$ and $a_w$ each denotes the forward cut and the backward cut as well as $m_{\rm max}$ represents the maximal trip completion derived from the stationary cut. It should be noted that the horizontal cuts may be inactive under certain rare occasions, such as short links with deficient offsets, indicating the forward and backward cuts intercept under the horizontal cut $m_{\rm max}$. The skewness-based fragility indicator and the approximation function developed in this work are indifferent to such cases with inactive horizontal cuts. The maximal vehicle accumulation, denoted as $n_{\rm max}$, is another indispensable parameter in MoC, and we fix $n_{\rm max}$ at $10000$ to create the unit MFD, roughly resembling the $n_{\rm max}$ generated for the city center of Zurich in the following Section \ref{sec: numerical}. An MFD for any given network can be scaled up or down to match the $n_{\rm max}$ of the proposed unit MFD without changing its original skewness, with a mathematical proof of such scalability provided in \ref{sec: scale}. The skewness of the network, denoted as $s$, can be computed as the asymmetry of the probability density distribution of TTS from $N_{sample}$ disruption recovery processes on the same MFD but under different disruption magnitudes. With $\mu$ and $\sigma$ each denoting the mean and the standard deviation, the skewness can be formulated as: 

\begin{align}
    \label{eq: skewness}
    s=\frac{1}{N_{sample}}\sum_{i=1}^{N_{sample}}\left(\frac{TTS_i-\mu}{\sigma}\right)^3
\end{align}

\begin{figure}[hbt!]
  \centering
    \subfigure[$m_{\rm max}=0.5$]{
      \hspace{-1cm}\resizebox*{7 cm}{!}{\adjincludegraphics[trim={0 0 {.18\width} 0},clip]
      {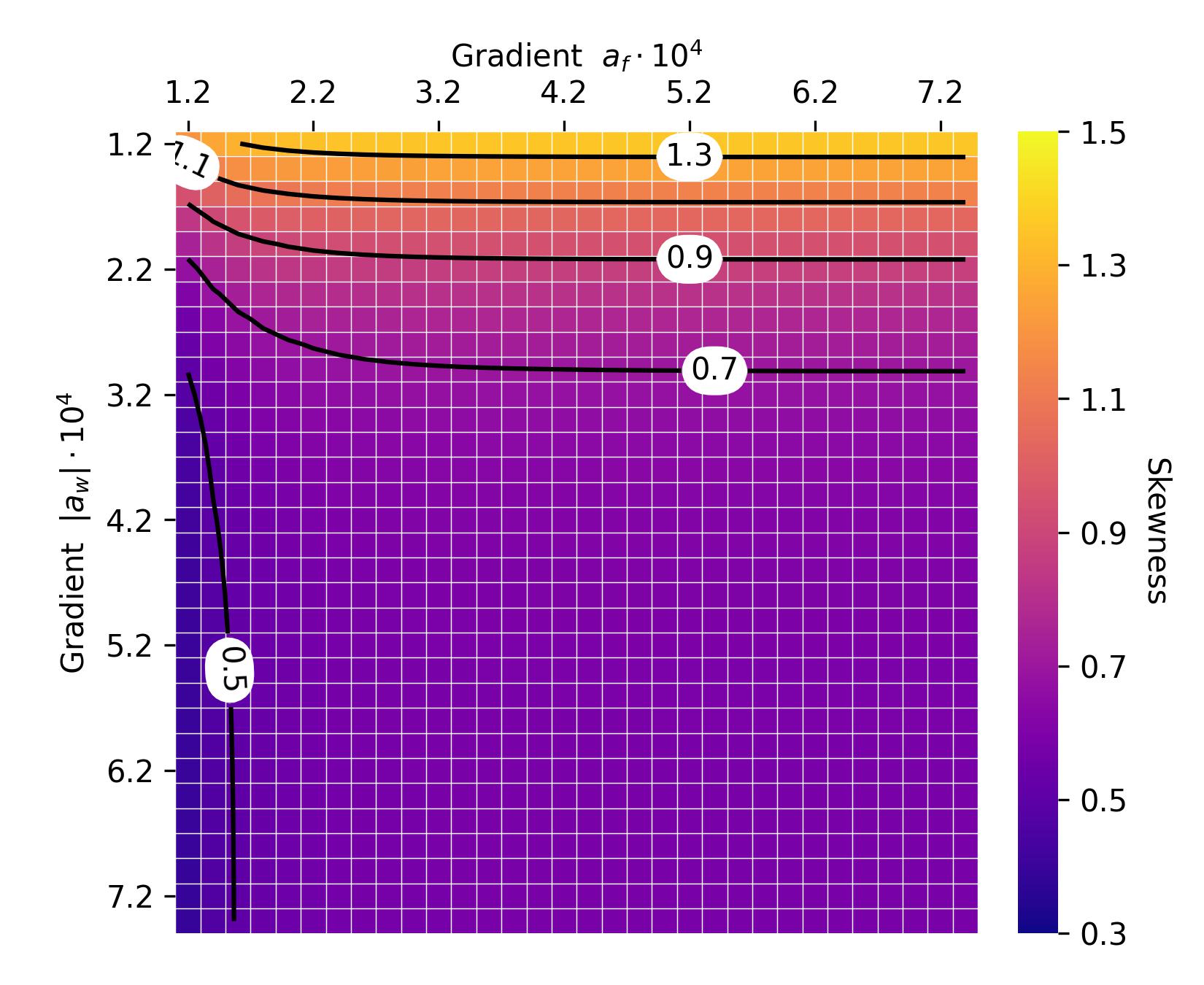}}
      \label{fig: q=0.5}}
    \subfigure[$m_{\rm max}=0.75$]{
      \hspace{1cm}\resizebox*{7.35 cm}{!}{\adjincludegraphics[trim={{.14\width} 0 0 0},clip]{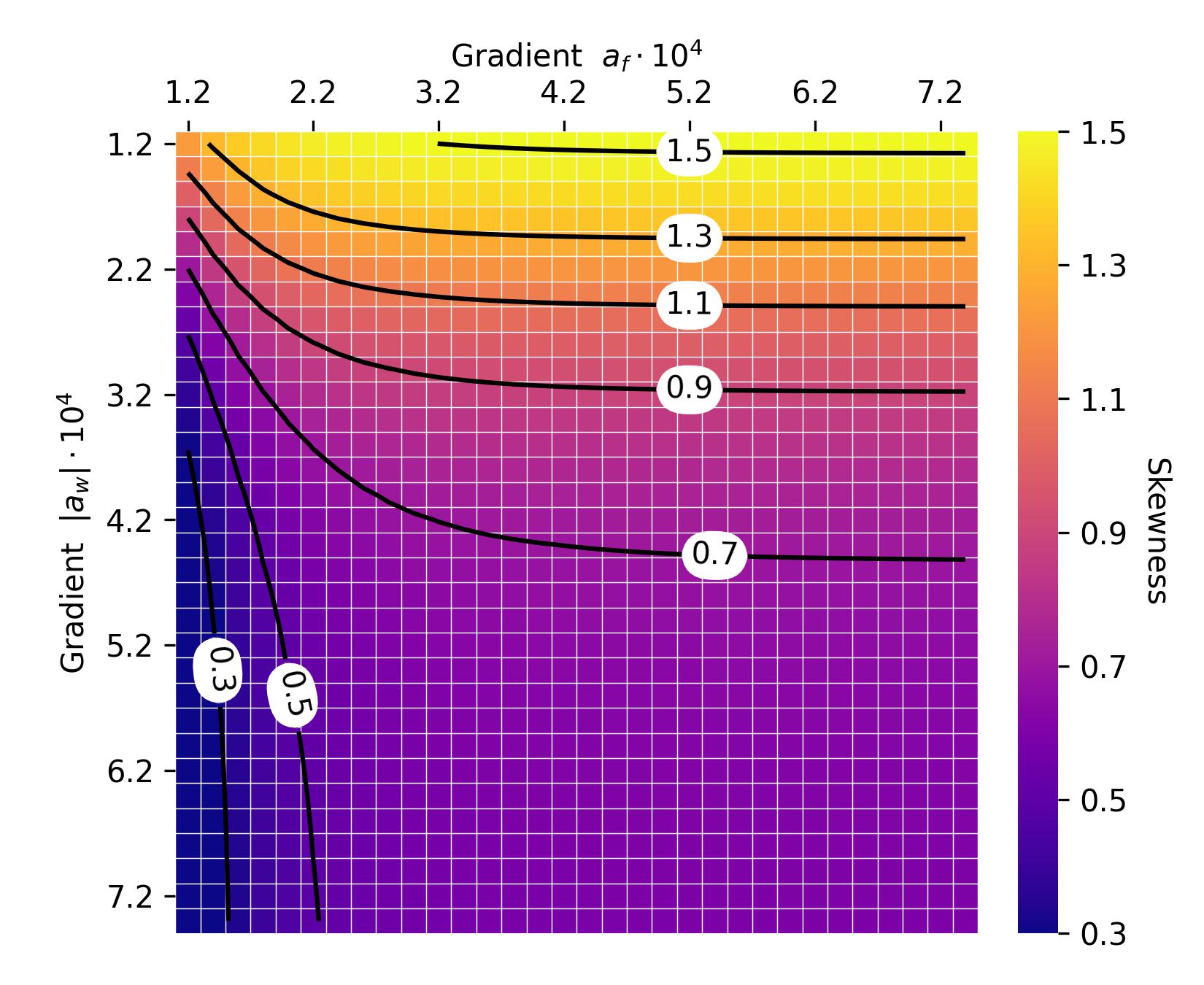}}
      \label{fig: q=0.75}}
    \subfigure[$m_{\rm max}=1.0$]{
     \hspace{-1cm}\resizebox*{7 cm}{!}{\adjincludegraphics[trim={0 0 {.18\width} {.1\width}},clip]{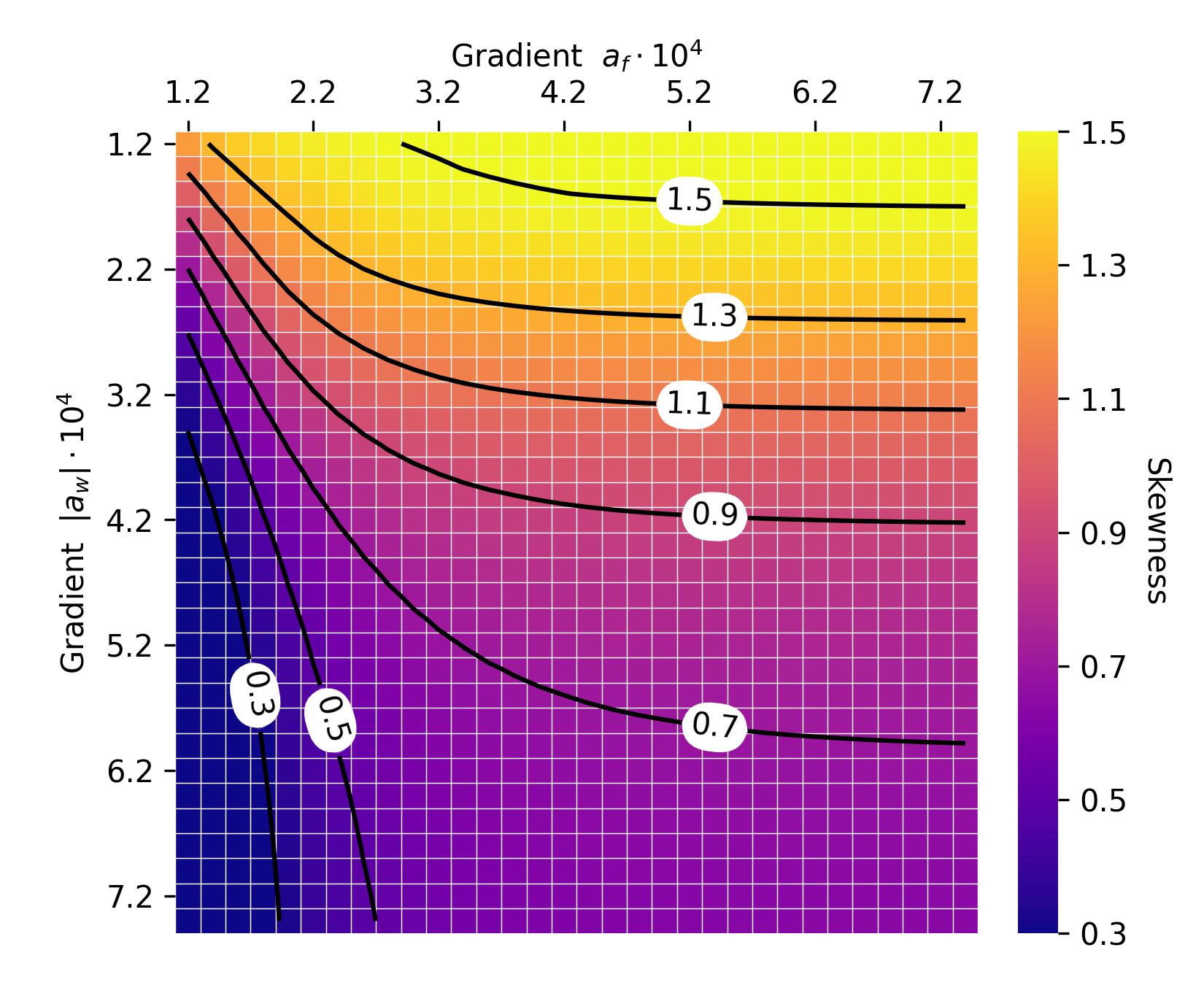}}
      \label{fig: q=1}}
    \subfigure[$m_{\rm max}=1.25$]{%
      \hspace{1cm}\resizebox*{7.35 cm}{!}{\adjincludegraphics[trim={{.14\width} 0 0 {.1\width}},clip]{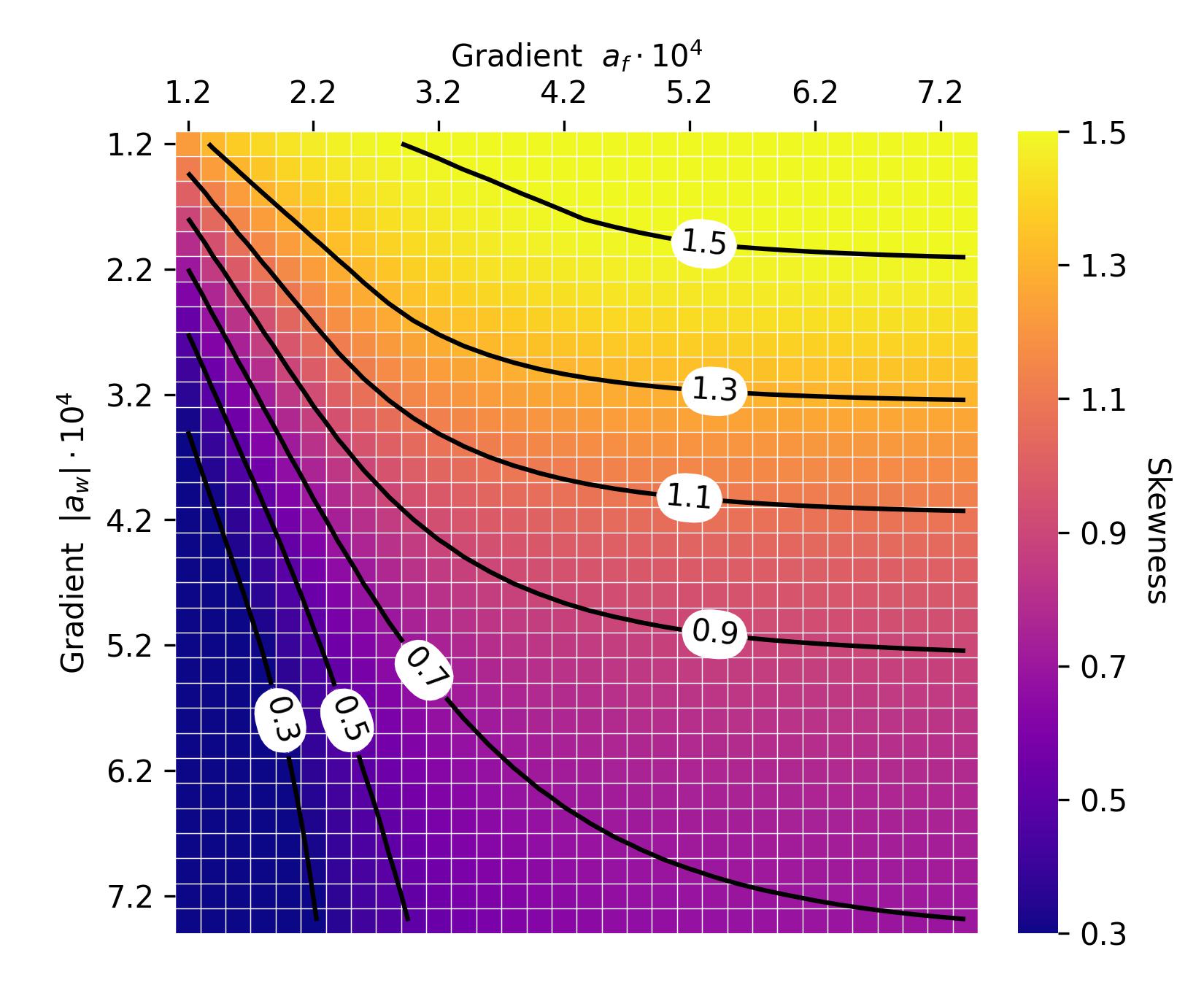}}
      \label{fig: q=1.25}}
    \subfigure[$m_{\rm max}=1.5$]{%
      \hspace{-1cm}\resizebox*{7 cm}{!}{\adjincludegraphics[trim={0 0 {.18\width} {.1\width}},clip]{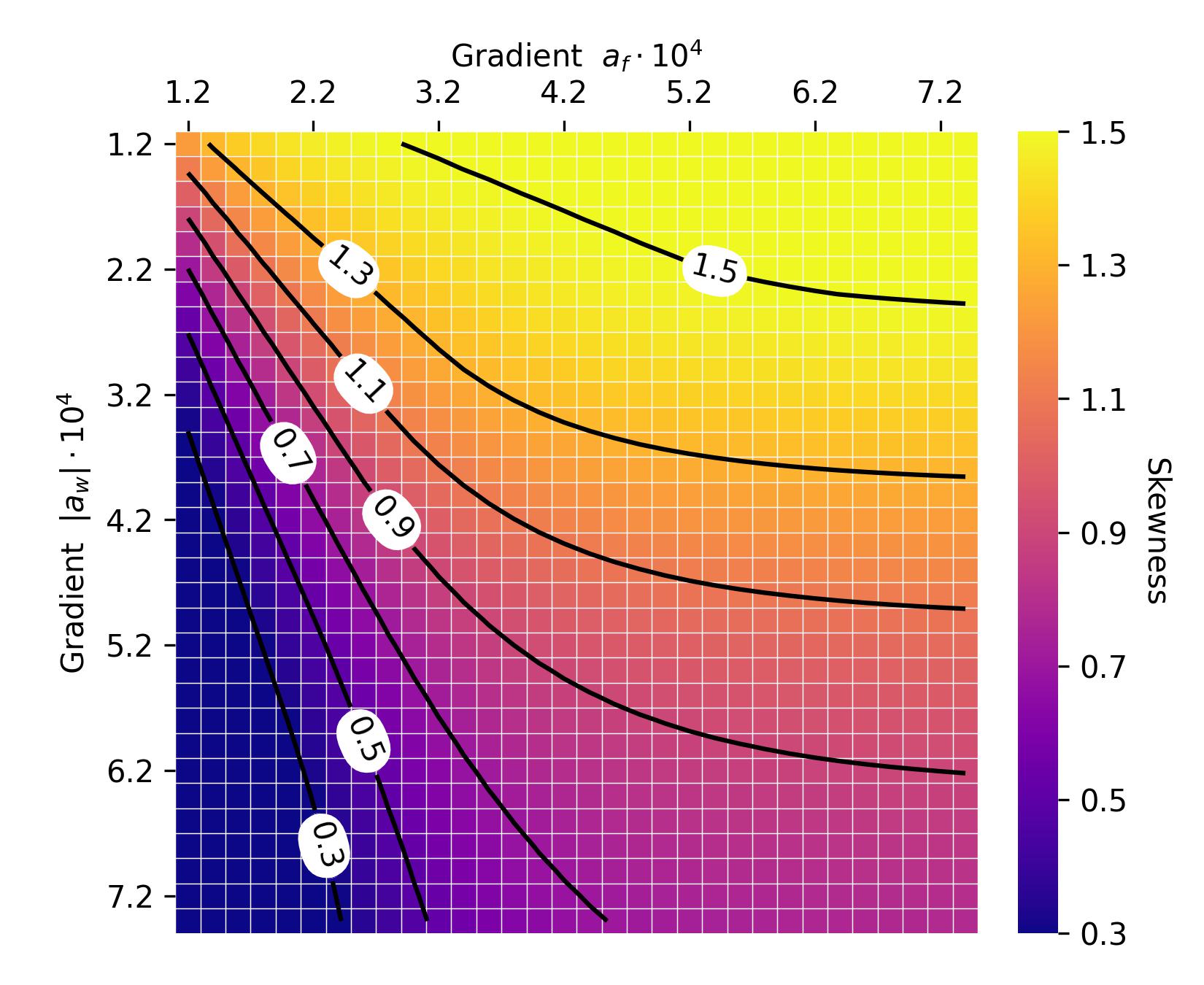}}
      \label{fig: q=1.5}}
    \subfigure[$m_{\rm max}=1.75$]{%
      \hspace{1cm}\resizebox*{7.35 cm}{!}{\adjincludegraphics[trim={{.14\width} 0 0 {.1\width}},clip]{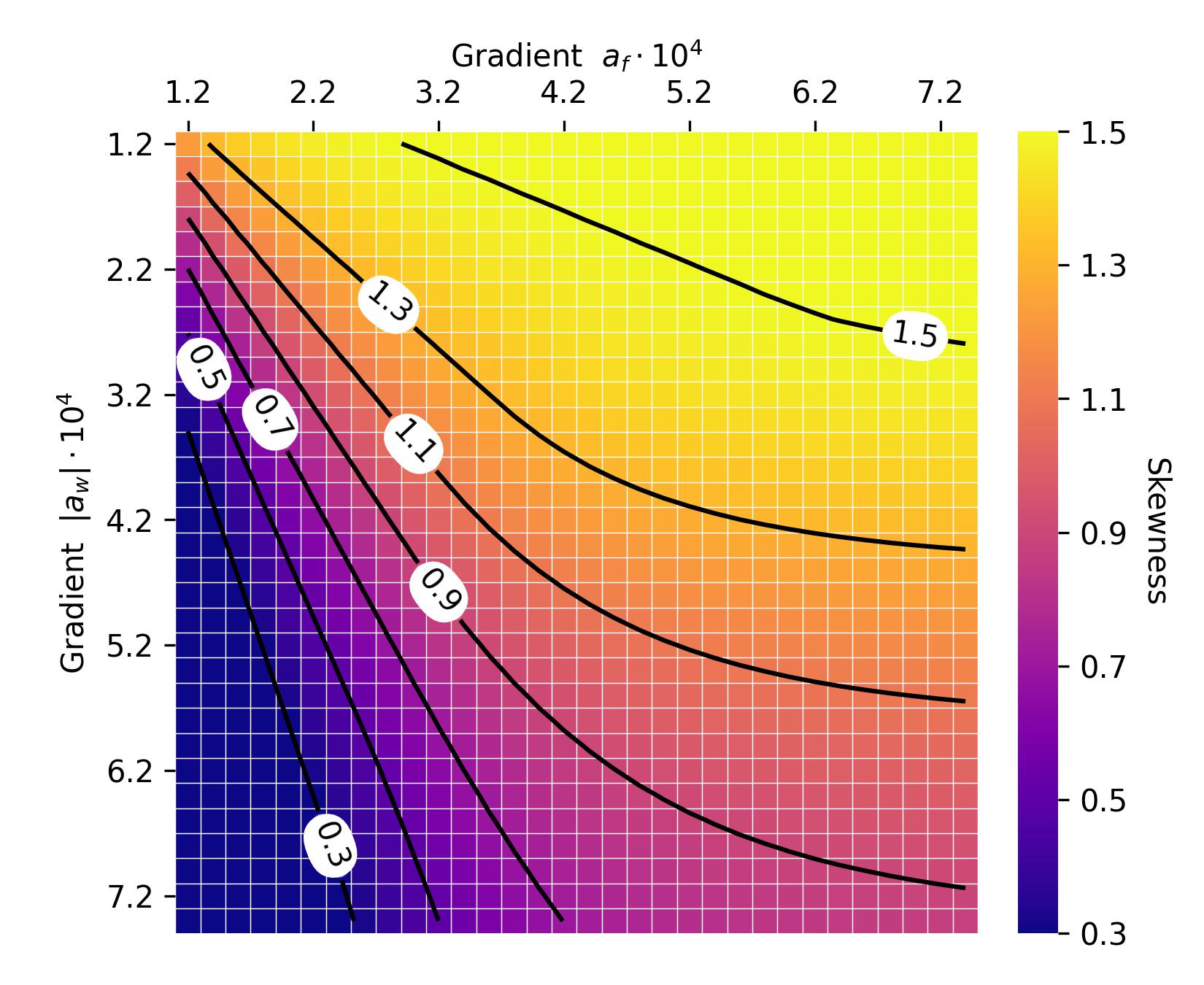}}
      \label{fig: q=1.75}}
    \caption{Skewness $s$ heatmap with $a_f$ and $a_w$ across different $m_{\rm max}$.} 
    \label{fig: heatmap}
\end{figure}

To ensure fair comparison, certain criteria must be kept consistent. First, it is assumed that there is no base demand in the network. Second, when calculating TTS, the initial disruption vehicle accumulation $n'$ follows a uniform distribution ranging between $5\%$ and $95\%$ of the maximal accumulation $n_{\rm max}$. The interval for vehicle increment for this setting is $50$, so a total number of $180$ data points are sampled. It is worth highlighting that to calculate skewness, a sufficient number of samples is required to yield satisfactory accuracy, and our chosen number of samples is substantially above the threshold based on pre-experiments. Also note these conditions merely serve as a synchronized reference rather than strict requirements, meaning that a range between $10\%$ and $90\%$ may also be used for determining the skewness. However, a larger range renders the skewness of different networks more distinguishable. Once the experiment setup is established, the aim is then to approximate the estimated skewness of a given network, denoted as $\tilde{s}$, based on its MFD-related parameters and a proposed approximation function $\tilde{s}=f_{\rm approx}(a_f, a_w, m_{\rm max})$. As per \cite{daganzo_cell_1994}, the backward wave travels several times slower than the free flow speed, so the FD is formulated with a wave speed $|w|\leq u_f$. We also adopt the same assumption $|a_w|\leq a_f$ for the development of the fragility indicator. In Fig. \ref{fig: heatmap}, we present the heatmaps of $s$ as a function of $|a_w|$ and $a_f$ for different values of $m_{\rm max}$. Both axes representing $a_f$ and $|a_w|$ begin from $1.2\cdot10^{-4}$ instead of $10^{-4}$ or even $0$, since with the maximal vehicle accumulation $n_{\rm max}=10000$, a gradient $a_f \leq\cdot10^{-4}$ will lead to exceedingly low $m_{\max}$. The contour lines are also illustrated to indicate multiple levels of skewness. Several patterns can be observed based on the contour lines, which represent an implicit function between skewness and the MFD-related parameters $|a_w|=f_{\rm approx}(a_f, m_{\rm max}, \tilde{s})$.

\begin{observation}
\label{obs: scalable}
\normalfont For the contour lines with the same skewness across different values of $m_{\rm max}$, it can be found that $m_{\rm max}$ has a scaling effect, which can be mathematically formulated as Eq. \ref{eq: scaling}. For example, in Fig. \ref{fig: q=1.5} with $m_{\rm max}=1.5$, the contour line with $s=1.3$ crosses approximately the point $(a_f, |a_w|)=(6.2, 3.8)$, which means in Fig. \ref{fig: q=0.75} with $m_{\rm max}=0.75$, half of the $m_{\rm max}$ in Fig. \ref{fig: q=1.5}, the same contour line $s=1.3$ will cross the point $(a_f, |a_w|)=(3.1, 1.9)$. This observation will be further discussed and validated in Section \ref{sec: validation}.
\end{observation}

\begin{align}
\label{eq: scaling}
\frac{|a_w|}{m_{\rm max}}=f_{\rm approx}\biggr(\frac{a_f}{m_{\rm max}}, \tilde{s}\biggr)
\end{align}

\begin{figure}[hbt]
  \centering
    \subfigure[Approximation of $\lim_{a_f\to\infty}|a_w|$]{%
      \resizebox*{8.cm}{!}{\includegraphics{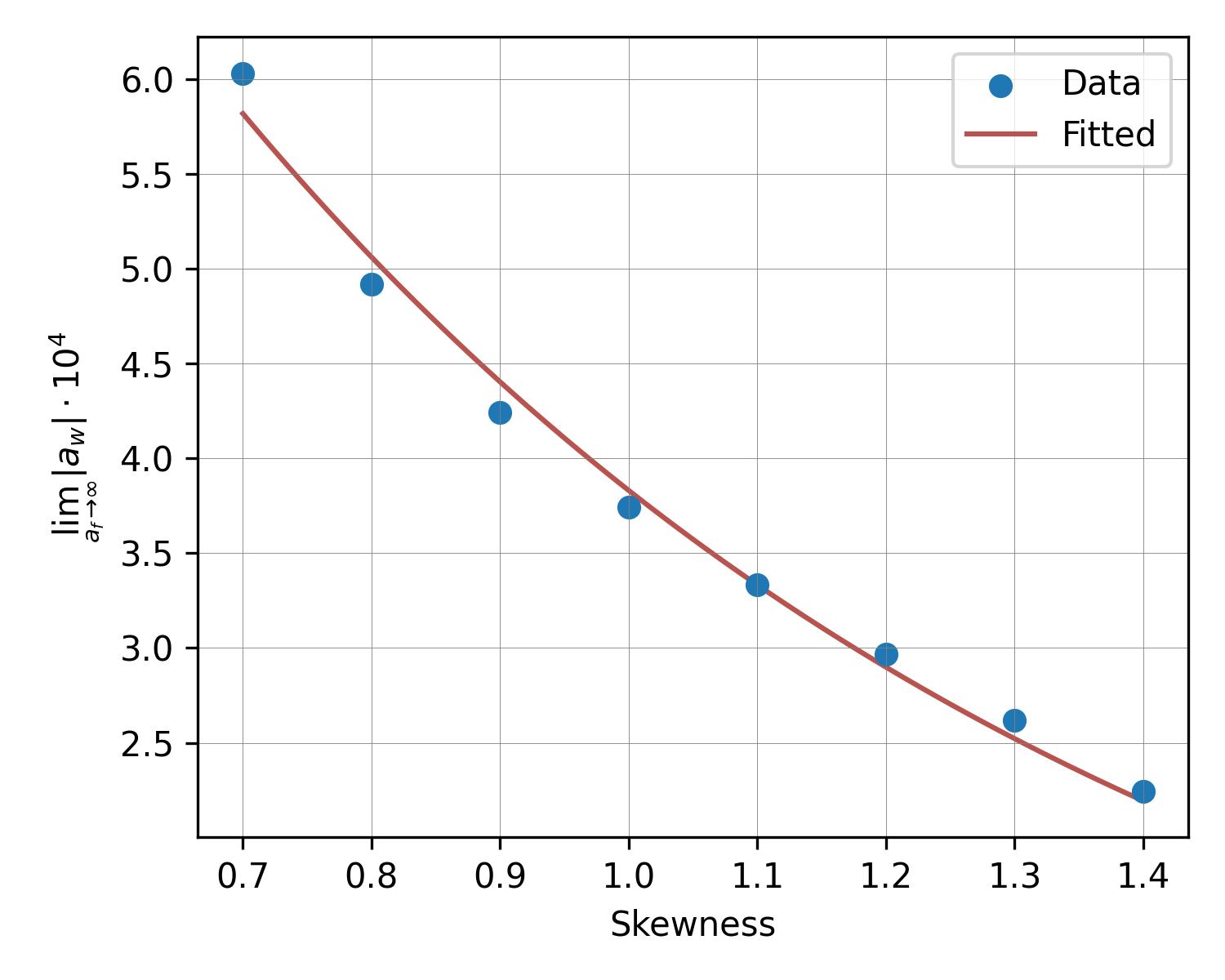}}\label{fig: c}}
      \hspace{10pt}
    \subfigure[Approximation of $\lim_{a_f\to0}|\frac{a_w}{a_f}|$]{%
      \resizebox*{8.cm}{!}{\includegraphics{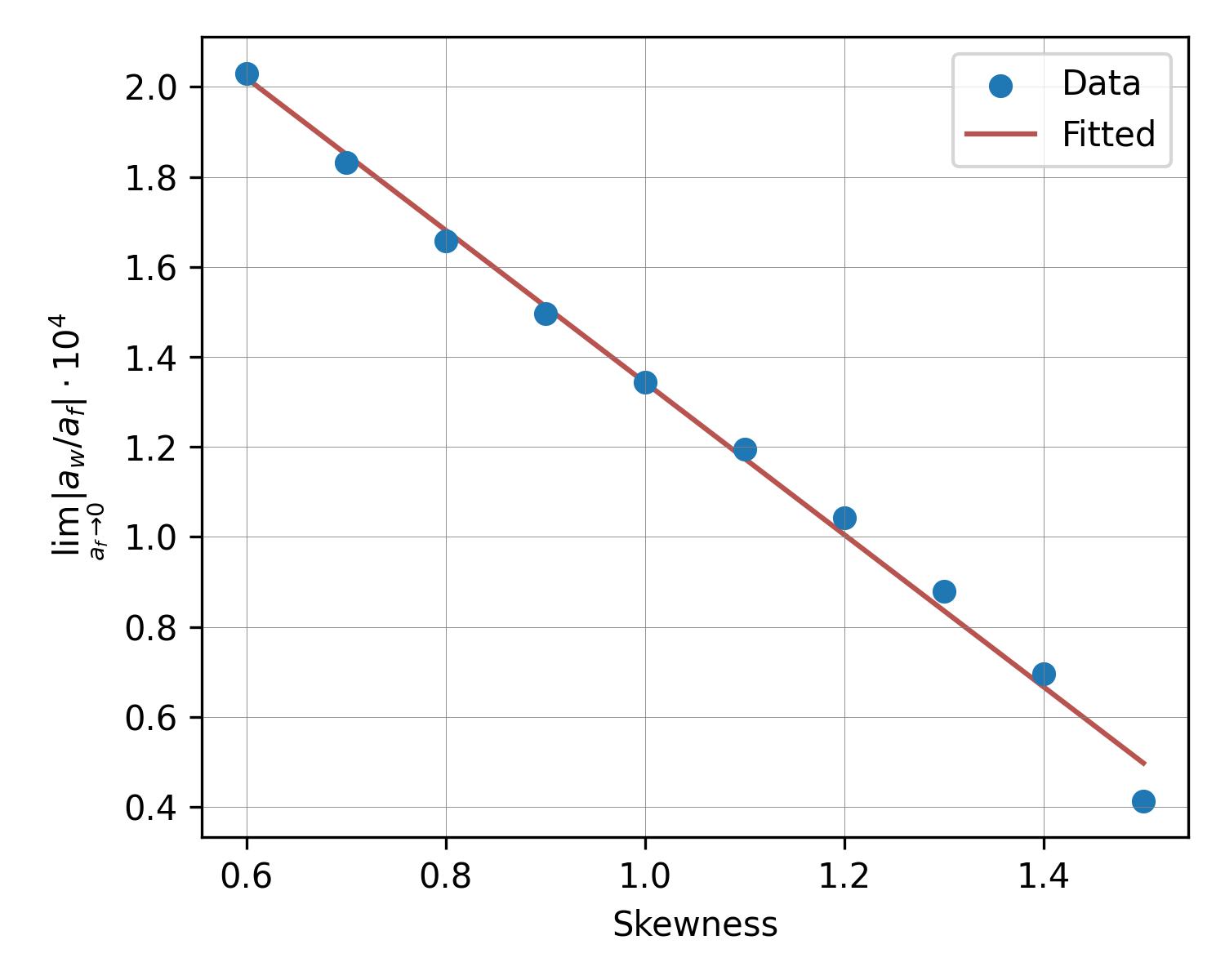}}\label{fig: k}}
    \caption{Approximation of the coefficients.} 
    \label{fig: approximation}
\end{figure}

\begin{observation}
\label{obs: constant}
\normalfont According to the assumption that $|a_w|\leq a_f$, we focus on the upper triangular matrix. And for each skewness contour line within the upper triangular matrix, $|a_w|$ becomes a constant value when $a_f$ increases to infinity, although the contour lines with very low skewness in the lower triangular matrix do not specifically follow this observation. An exponential relationship can be roughly approximated between $|a_w|$ and skewness $\tilde{s}$ with coefficients $\beta_1$, $\beta_2$, and $\beta_3$, as shown in Fig. \ref{fig: c}, which can be mathematically formulated as:
\end{observation}

\begin{align}
\label{eq: constant}
\lim_{a_f\to\infty}|a_w| = \beta_1 e^{\beta_2(\tilde{s}-\beta_3)}
\end{align}

\begin{observation}
\label{obs: linear}
\normalfont For each skewness contour line, when $a_f$ gets closer to $0$, the initial gradient of the contour line $\lim_{a_f\to0}|\frac{a_w}{a_f}|$ will also become a constant. And when extended towards the upper left, they intercept exactly at the origin of the coordinates $(a_f, |a_w|)=(0, 0)$. The relationship between the gradient of the contour line $\lim_{a_f\to0}|\frac{a_w}{a_f}|$ and skewness $\tilde{s}$ can be roughly approximated as a linear function, as shown in Fig. \ref{fig: k}, which can be mathematically formulated as:
\end{observation}

\begin{align}
\label{eq: linear}
\lim_{a_f\to0}|\frac{a_w}{a_f}| = \beta_4 \tilde{s} + \beta_5
\end{align}

\subsection{Skewness-based fragility indicator}
The above observations suggest a strong correlation between the skewness and the MFD-related parameters worth further exploitation. Based on Observation \ref{obs: constant} and \ref{obs: linear}, these contour lines have been found to share great similarities with certain types of functions in the family of sigmoid curves. We refer such approximators to as activation functions $f_{\rm act}(x)$, with the term borrowed from the deep learning domain, where the hyperbolic tangent $f_{\tanh}(x)=\frac{e^x-e^{-x}}{e^x+e^{-x}}$ is one of the most commonly applied functional forms. Similar to Observation \ref{obs: constant}, the hyperbolic tangent function has a constant value at the right end $\lim_{x\to\infty}f_{\tanh}(x)=1$. Another comparable property can be observed is that the gradient of the activation function at the origin is also a constant $f'_{\tanh}(0)=1$ as Observation \ref{obs: linear}.

To scale the activation function to a certain skewness contour line according to these two observations to generate our proposed approximation function $f_{\rm approx}(\cdot)$, the following two conditions are to be satisfied: 

\begin{subequations}
\begin{align}
    \lim_{a_f\to\infty}f_{\rm approx}(a_f) = \biggr(\lim_{a_f\to\infty}|a_w|\biggr) \cdot \biggr(\lim_{a_f\to\infty}f_{\tanh}(a_f)\biggr)
    \\
    \lim_{a_f\to0}\frac{f_{\rm approx}(a_f)}{a_f} = \biggr(\lim_{a_f\to0}|\frac{a_w}{a_f}|\biggr) \cdot \biggr(\lim_{a_f\to0}\frac{f_{\tanh}(a_f)}{a_f}\biggr)
\end{align}
\end{subequations}

Therefore, the hyperbolic tangent $f_{\tanh}(x)$ for the approximation of skewness $\tilde{s}$ needs to be rewritten as:

\begin{align}
\nonumber
|a_w| & = f_{\rm approx} (a_f)
\\
& = \lim_{a_f\to\infty}|a_w| \cdot f_{\tanh} \biggr(\frac{\lim_{a_f\to0}|\frac{a_w}{a_f}|}{\lim_{a_f\to\infty}|a_w|}\cdot a_f\biggr)
\end{align}

Now taking Observation \ref{obs: scalable} into account as well, and with coefficients $\beta_{1,...,5}$ being able to be approximated with Eq. \ref{eq: constant}, Eq. \ref{eq: linear}, and Fig. \ref{fig: approximation}, the proposed approximation function is:

\begin{align}
\label{eq: activation}
\frac{|a_w|}{m_{\rm max}} = \beta_1 e^{\beta_2(\tilde{s}-\beta_3)} \cdot f_{\tanh} \biggr(\frac{\beta_4 \tilde{s} + \beta_5}{\beta_1 e^{\beta_2(\tilde{s}-\beta_3)}}\cdot \frac{a_f}{m_{\rm max}}\biggr)
\end{align}

Despite being an implicit function, the skewness $\tilde{s}$ can be solved with an iterative optimization algorithm, such as the Powell hybrid method and its variations. With both ends $a_f\rightarrow0$ and $a_f\rightarrow\infty$ of the contour lines being fixed, the hyperbolic tangent function alone does not necessarily guarantee a sufficiently decent approximation. Hence, we introduce other activation functions that resemble the hyperbolic tangent function in a list below, which will be applied and cross-compared for the approximation of the skewness of a network. All these activation functions share the same characteristics that $\lim_{x\to\infty}f_{\rm act}(x)=1$ and $f'_{\rm act}(0)=1$, but with varying curvature and trajectory to reach $a_f\rightarrow\infty$:

\begin{itemize}
    \item Error function: $f_{\rm erf}(x)= \frac{2}{\sqrt{\pi}}\int_0^x e^{-t^2}dt$
    \item Gudermannian function: $f_{\rm gd}(x)=2 \arctan(\tanh(\frac{x}{2}))$
    \item Arctangent function: $f_{\rm arctan}(x)=\frac{2}{\pi}\arctan(\frac{2}{\pi} x)$
    \item Inverted Square Root Unit (ISRU): $f_{\rm isru}(x)=\frac{x}{\sqrt{1+x^2}}$
    \item Generalized form of ISRU:  $f_{\kappa}(x)=\frac{x}{(1+|x|^\kappa)^{1/\kappa}}$  
\end{itemize}

\subsection{Validation}
\label{sec: validation}
\begin{figure}[hbt!]
  \centering
    \subfigure[$f_{\kappa=4}$]{
      \resizebox*{6.6 cm}{!}{\adjincludegraphics[trim={0 0 {.20\width} 0},clip]{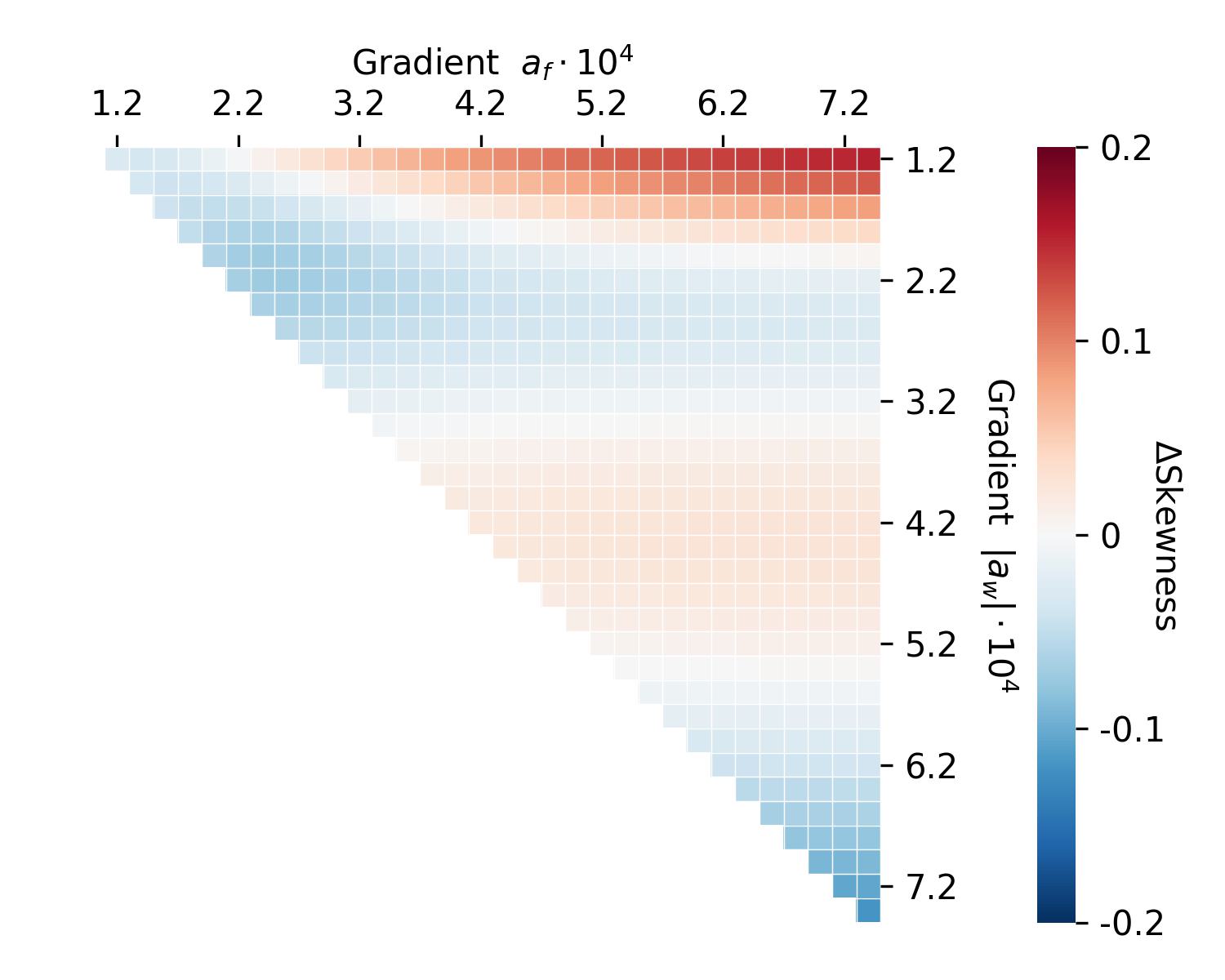}}
      \label{fig: i4ru}}
    \subfigure[$f_{\rm erf}$]{
      \hspace{1cm}\resizebox*{8.2 cm}{!}{\adjincludegraphics[trim={0 0 0 0},clip]{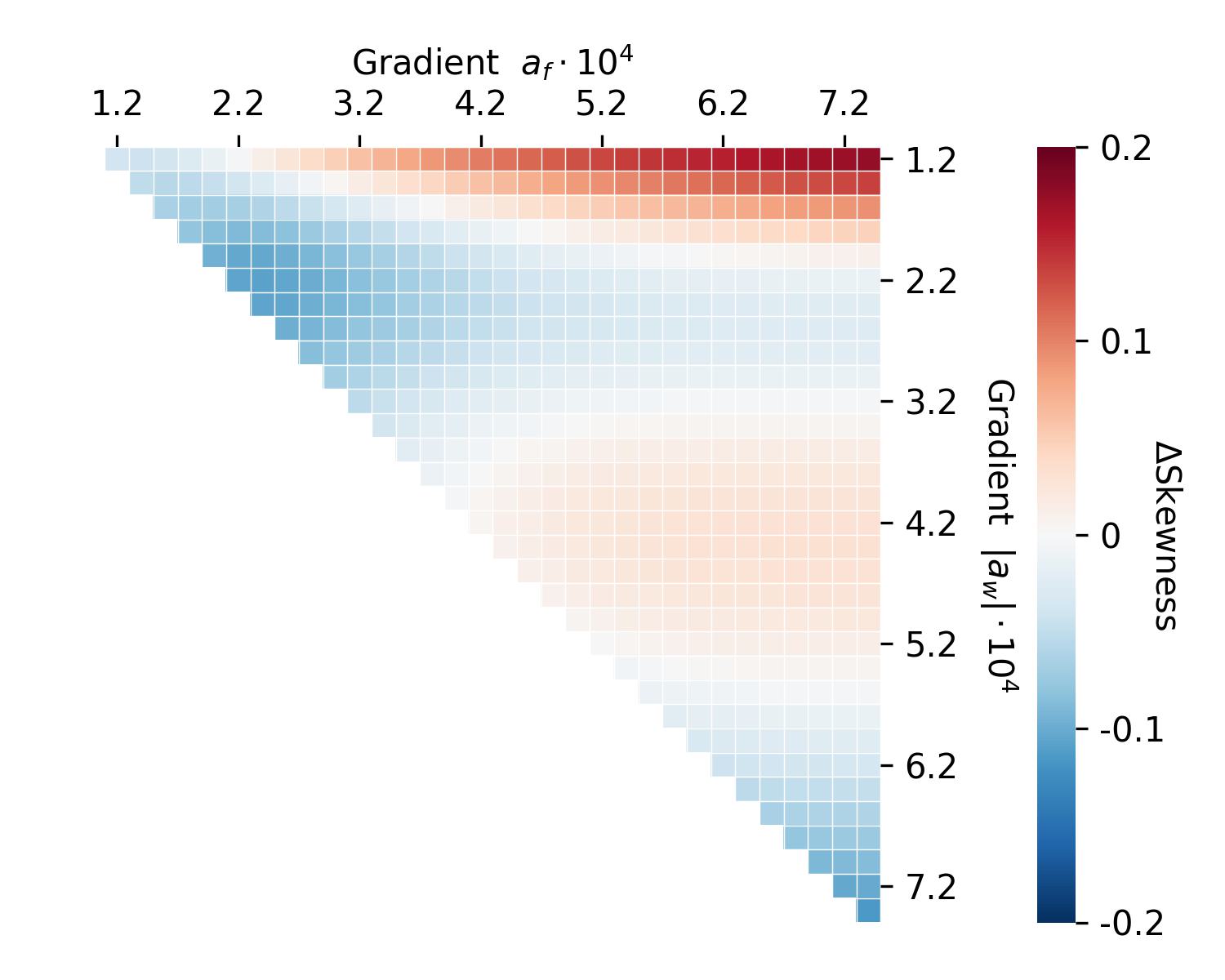}}
      \label{fig: erf}}
    \subfigure[$f_{\rm tanh}$]{
      \resizebox*{6.6 cm}{!}{\adjincludegraphics[trim={0 0 {.20\width} {.065\width}},clip]{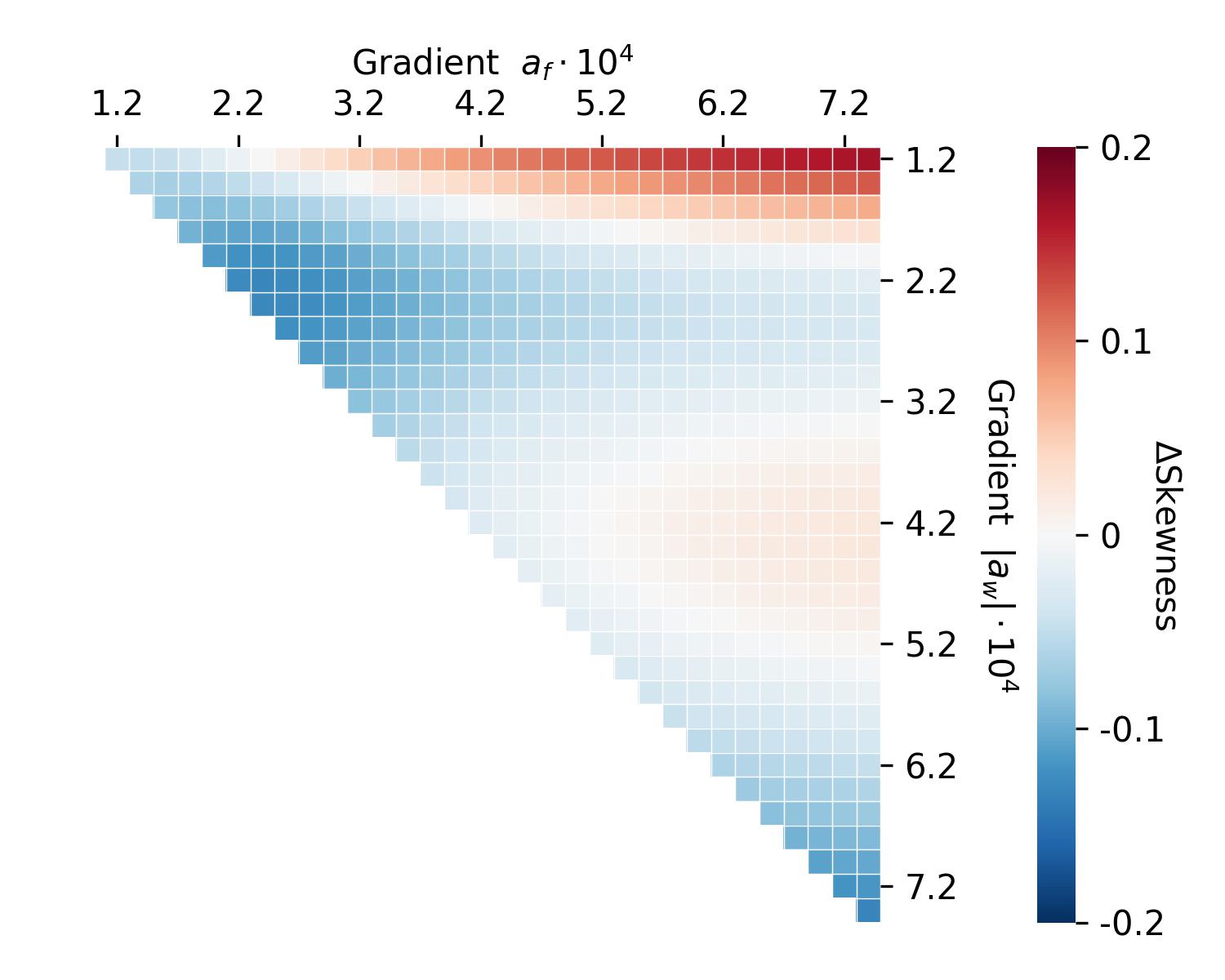}}
      \label{fig: tanh}}
    \subfigure[$f_{\rm gd}$]{%
      \hspace{1cm}\resizebox*{8.2 cm}{!}{\adjincludegraphics[trim={0 0 0 {.065\width}},clip]{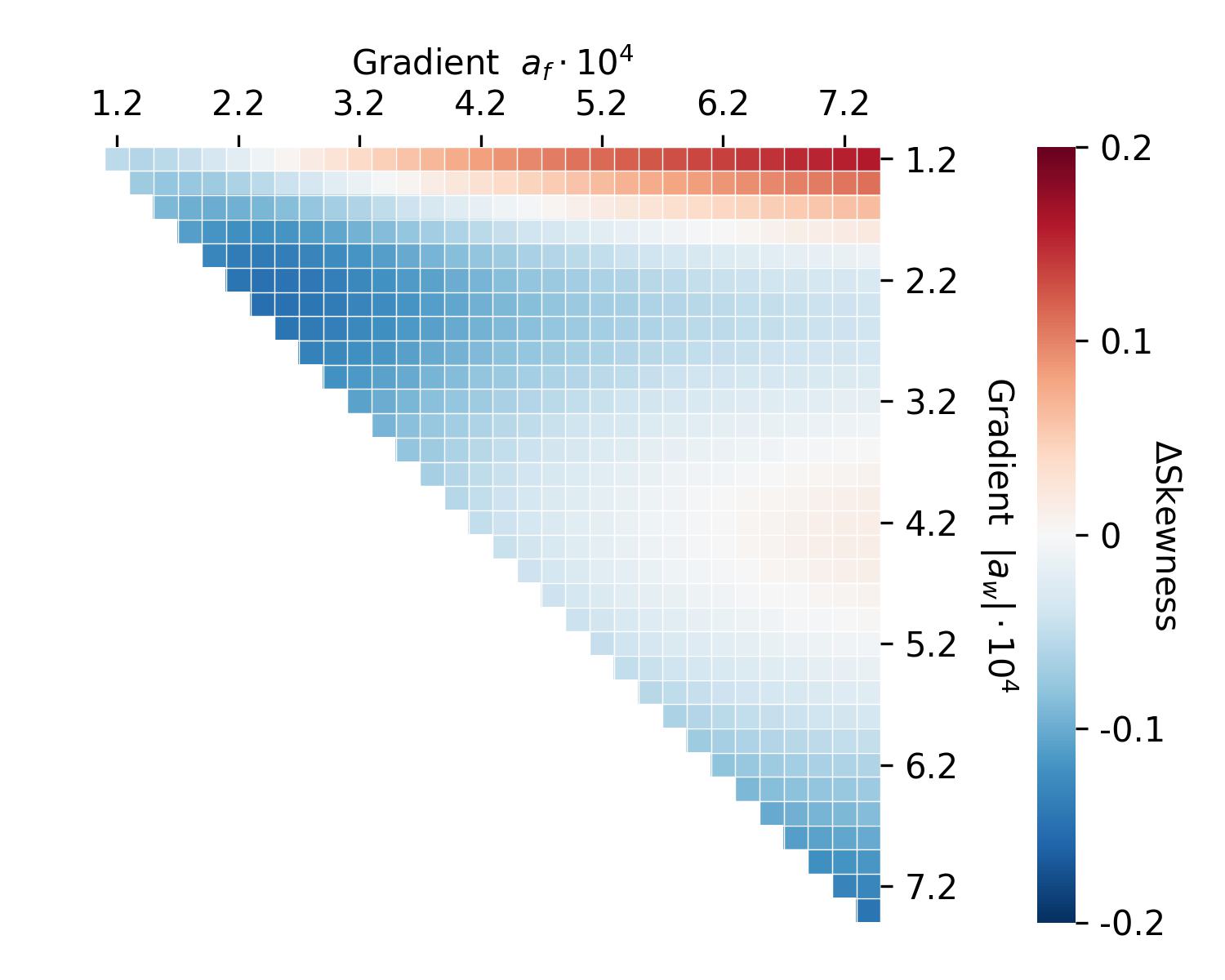}}
      \label{fig: gd}}
    \subfigure[$f_{\rm isru}$]{%
      \resizebox*{6.6 cm}{!}{\adjincludegraphics[trim={0 0 {.20\width} {.065\width}},clip]{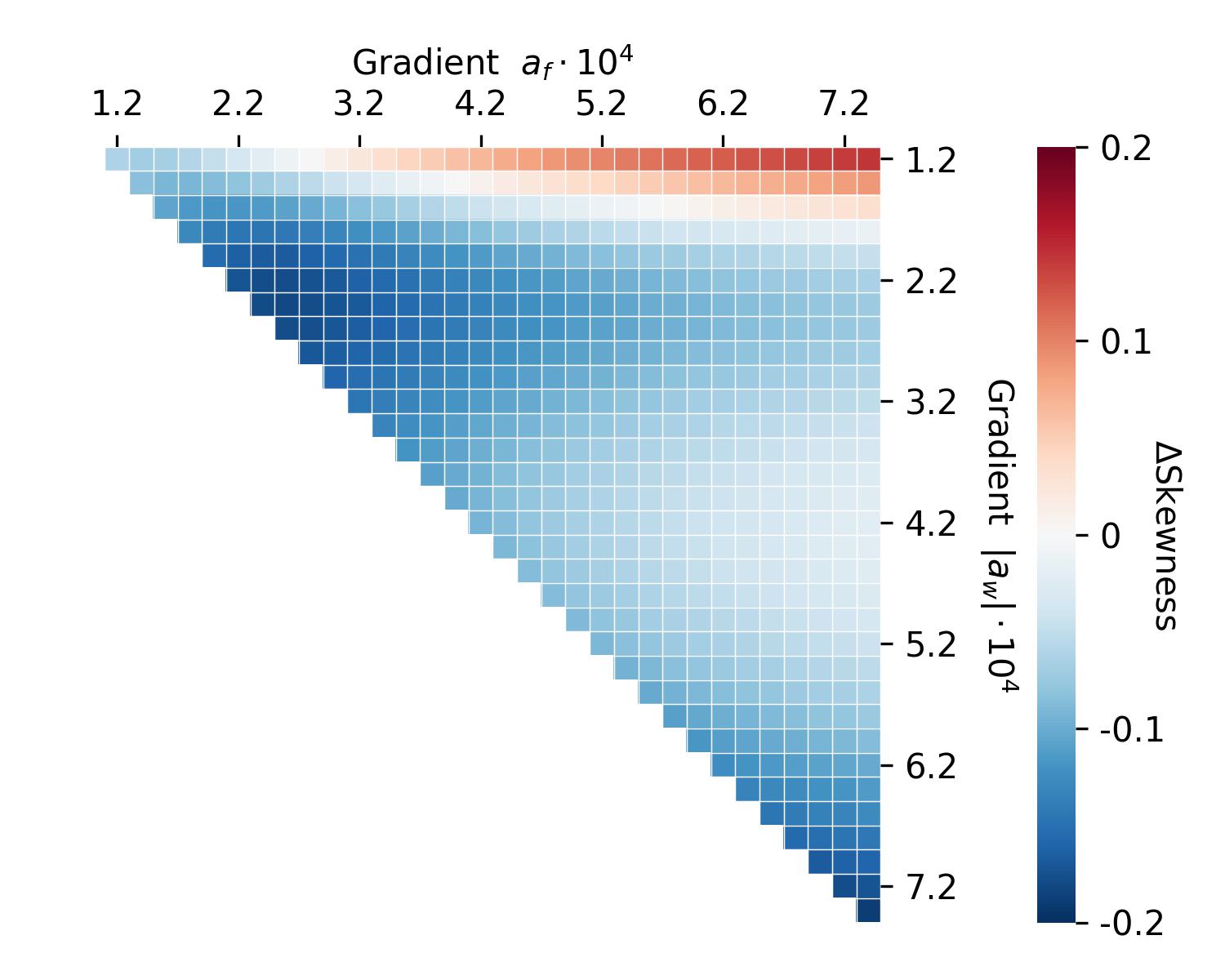}}
      \label{fig: isru}}
    \subfigure[$f_{\rm arctan}$]{%
      \hspace{1cm}\resizebox*{8.2 cm}{!}{\adjincludegraphics[trim={0 0 0 {.065\width}},clip]{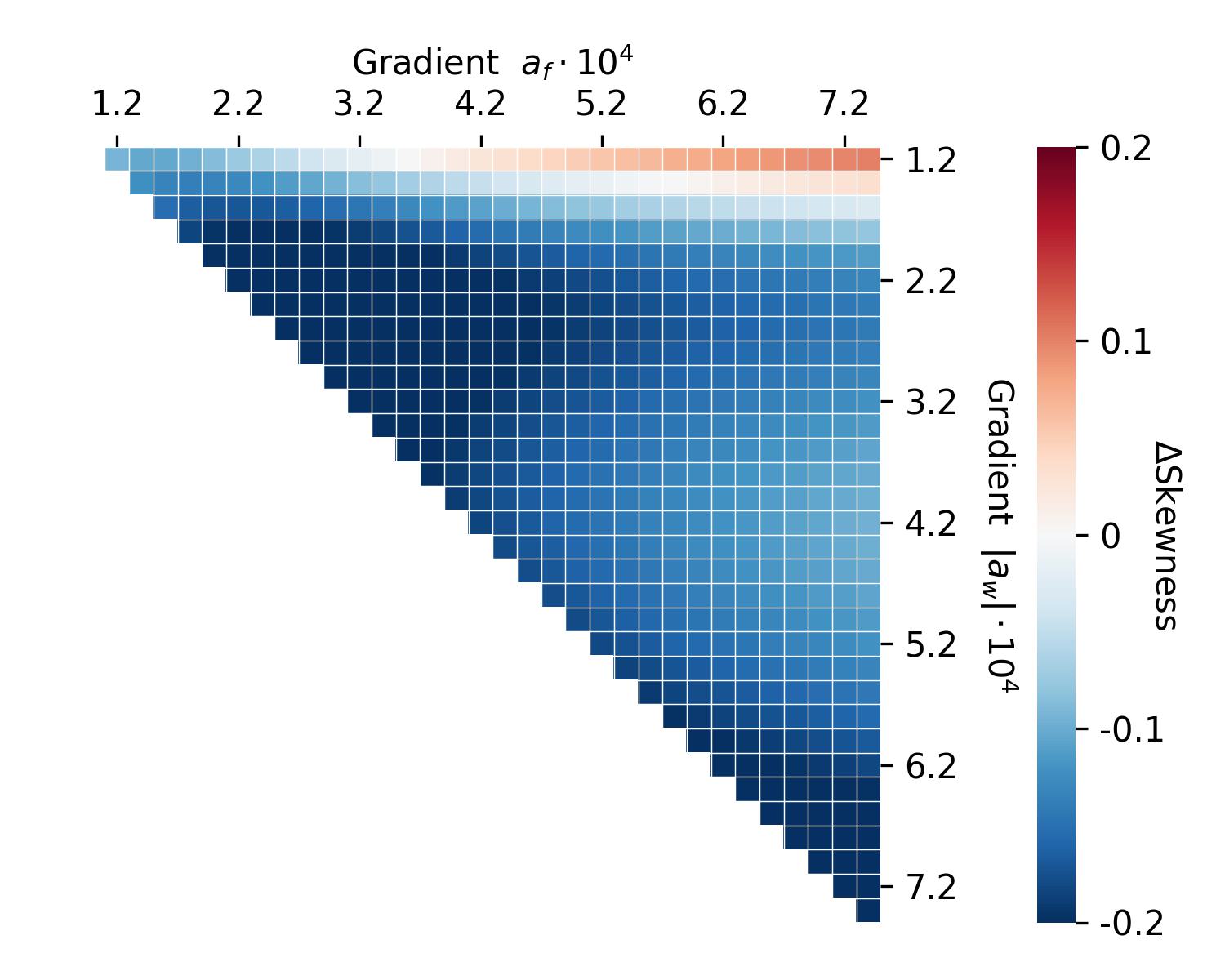}}
      \label{fig: arctan}}
    \caption{Error of approximation by applying different activation functions with $m_{\rm max}=1$.} 
    \label{fig: error}
\end{figure}

To validate the effectiveness of our proposed skewness-based fragility indicator, the approximation error $\Delta s$ between the real skewness $s$ from the unit MFD and the approximated skewness $\tilde{s}$ of each activation function is computed and presented in Fig. \ref{fig: error} with $m_{\rm max}=1$. In Fig. \ref{fig: i4ru} we select $k=4$ to represent the generalized form of the inverted square root unit. A darker shape of red or blue indicates the skewness from the approximation $\tilde{s}$ is greater or smaller than the real value $s$. It can be observed that the activation functions $f_{\kappa=4}$ and $f_{\rm erf}$ show seemingly superior performance, with the vast majority of the heatmap cells lying below an error of $0.05$. It should also be noted that at the upper margin of the heatmap with $a_f \gg |a_w|$, a red strip with significant positive error can be observed. We acknowledge but accept such deviation since this red strip with $a_f \gg |a_w|$ lacks realism in a way similar to the lower triangular matrix with $a_f<|a_w|$. 

To better quantify the error $\Delta s$ from selected approximation functions, we summarize the error in Table \ref{tab: error} with Mean Absolute Error (MAE), Mean Squared Error (MSE), and Root Mean Square Error (RMSE). Selected values of $\kappa$ for the generalized form of the inverted square root unit $f_{\kappa}$ are also presented. The activation function $f_{\kappa=4}$ showcased the best performance, with MAE and RMSE being $0.033$ and $0.043$ respectively. Considering the skewness in the upper triangular matrix in Fig. \ref{fig: q=1} varies roughly between $0.7-1.5$ and shows an average skewness of $1.136$, the MAE and RMSE demonstrate an error well below 5\% and thus validate the accuracy of our proposed approximation function. With $f_{\kappa=4}$, we also showcase the approximation accuracy for varying values of $m_{\rm max}$ in Table \ref{tab: error}. While the error still lies below $5\%$ with $m_{\rm max}$ larger than $1$, the accuracy deteriorates when $m_{\rm max}$ becomes smaller. This is partially attributable to the function form of approximation in Eq. \ref{eq: constant} and Eq. \ref{eq: linear} as well as the coefficients $\beta_{1,...,5}$. The current coefficients are approximated with the data points from a trimmed range in Fig. \ref{fig: approximation} to achieve the least error in the case of $m_{\rm max}=1$. If a slightly larger range is considered, a rather uniform value of RMSE at around $0.75-0.85$ can be observed regardless of the different values of $m_{\rm max}$, roughly $7\%$ of the average skewness, validating Observation \ref{obs: scalable}. Also, for the approximation of $\lim_{a_f\to\infty}|a_w|$, a polynomial can be applied in place of the current exponential function to achieve potentially better performance. Note that the error computation always accounts for the red strip on the top, and can be further reduced if it is excluded due to its lack of realism.

\begin{table}[h]
\small
\centering
\renewcommand{\arraystretch}{1.5}
\caption{Error of approximation of applying different activation functions.}
\label{tab: error}
\begin{tabular}{C{2.2cm}C{1.5cm}C{1.5cm}C{1.5cm}|C{2.2cm}C{1.5cm}C{1.5cm}C{1.5cm}}
\toprule
\textbf{Function} & MAE & MSE & RMSE & \textbf{With $f_{\kappa=4}$} & MAE & MSE & RMSE 
\\ \midrule
$f_{\rm erf}$   & 0.037 & 0.0023 & 0.048 & $m_{\rm max}=0.50$ & 0.121 & 0.035 & 0.187
\\ 
$f_{\rm tanh}$  & 0.044 & 0.0033 & 0.057 & $m_{\rm max}=0.75$ & 0.045 & 0.0047 & 0.068
\\
$f_{\rm gd}$    & 0.054 & 0.0045 & 0.067 & $m_{\rm max}=1.00$ & 0.033 & 0.0019 & 0.043  
\\
$f_{\rm arctan}$ & 0.152 & 0.0261 & 0.162 & $m_{\rm max}=1.25$ & 0.036 & 0.0023 & 0.048
\\
$f_{\kappa=2 \ (\rm isru)}$  & 0.084 & 0.0089 & 0.094 & $m_{\rm max}=1.50$ & 0.042 & 0.0029 & 0.054  
\\
$f_{\kappa=4}$  & \textbf{0.033} & \textbf{0.0019} & \textbf{0.043} & $m_{\rm max}=1.75$ & 0.046 & 0.0033 & 0.057
\\
$f_{\kappa=6}$  & 0.032 & 0.0020 & 0.044  &&&
\\
$f_{\kappa=8}$  & 0.033 & 0.0021 & 0.046 &&&
\\\bottomrule
\end{tabular}
\end{table}

\subsection{Implications}
Aside from the proposed approximation function to estimate the skewness-based fragility indicator, insights into how the shape of MFD affects the network fragility can also be drawn from the heatmaps in Fig. \ref{fig: heatmap}. When $|a_w|$ is not sufficiently large, an increase in $a_f$ contributes to the network's fragility, whereas a larger $|a_w|$ reduces such fragility. The implications are twofold: 
\begin{itemize}
    \item[-] To minimize urban traffic accident rates and reduce noise pollution, aligned with the long-term goal of Vision Zero in the EU and worldwide to reach minimal traffic fatalities and severe injuries, many cities have implemented stricter speed limits, such as the Tempo 30 regulation in Zurich \citep{menendez_implementing_2022}. Although this may reduce the overall serviceability of the urban road networks, it enhances the city's antifragility when dealing with traffic disruptions by decreasing $a_f$.
    \item[-] While backward wave speed is generally linked to the driving behavior of human-driven vehicles, \cite{makridis_platoon_2024} recently found that adaptive cruise control, a precursor to AVs, can increase backward wave speed with minimum headway settings (although still not as fast as free-flow speed), suggesting that introducing AVs into transportation systems could potentially improve the network's antifragility by increasing $|a_w|$. However, this enhancement of antifragility at the macroscopic level may well result in fragility at the microscopic level, as minimum safety distances can cause propagating string instability under perturbations \citep{makridis_openacc_2021}, if the connectivity and cooperation levels are not sufficiently high.
\end{itemize}

\section{Numerical simulation}
\label{sec: numerical}

Even though researchers generally consider MFDs to be well-defined, road transportation networks in the real world and their MFDs are always subject to stochasticity, as shown in \cite{geroliminis_existence_2008, ambuhl_disentangling_2021}. Therefore, when validating a newly proposed traffic control algorithm, it has become a common practice to account for model uncertainties and showcase the method's robustness, as in \cite{geroliminis_optimal_2013, zhou_scalable_2023}. In our study, however, the model stochasticity cannot be directly reflected in the mathematical analysis. Hence, it is indispensable to illustrate the influence of realistic stochasticity on the fragile nature of transportation networks with a numerical simulation, i.e., whether the system still maintains the same fragile response under real-world uncertainties in the MFD when a demand or supply disruption is present. 

\subsection{Experimental setup with real-world parameters}

In this section, we again simulate the traffic performance under the disruption recovery process. The MFD of the studied region is generated by applying MoC with real-world parameters based on the city center of Zurich. Some parameters, such as free-flow speed, back-propagation speed, maximal density, and capacity, are provided in \cite{ambuhl_functional_2020} for Zurich with queried routes from the Google API and with other validation methods. The total and average lane length for the city center is determined from the OpenStreetMap API. The average trip length of Zurich is studied in \cite{schussler2008identifying}. We introduce stochasticity in the city center of Zurich following real traffic light data, which is publicized by the Statistical Office of Zurich and accessible in \cite{genser_traffic_2023, genser_time--green_2024}. The authors acknowledge that MoC is developed with the premise of a homogeneous region, and given that the available data is limited to only one main intersection in this region, we assume that this intersection serves as a representative sample for the city center region. Moreover, since the vast majority of signalization in Zurich is actuated depending on the traffic flow \citep{RIEDEL201999}, they do not strictly follow a fixed-time signal cycle. Despite this actuation, a concentrated distribution can be observed in the dataset, and we assume the green split of the cycle follows a normal distribution. As the offset is presumed to be zero for Yokohama's responsive traffic light control in \cite{daganzo_analytical_2008}, the same assumption is applied in the Zurich network for its traffic signal actuation. According to the daily average traffic density of Zurich in \cite{ambuhl_disentangling_2021}, we approximate the traffic demand, which is also the trip completion when the traffic state is at equilibrium, as approximately $0.6$ veh/s for our studied region. This corresponds to a base accumulation of around 975 vehicles in the city center. The parameters and the related values are summarized in Table \ref{tab: parameters}.

\begin{table}[hbt]
\centering
\caption{Real-world parameters for the city center of Zurich.}
\label{tab: parameters}
\begin{tabular}{L{0.3\textwidth}C{0.1\textwidth}C{0.1\textwidth}C{0.1\textwidth}}
\toprule
Parameters             & Notation & Unit        & Value    
\\\midrule
Free-flow speed        & $u_l$    & $\rm m/s$   & 12.5     
\\
Backward wave speed    & $w_l$    & $\rm m/s$   & 6.0      
\\
Maximal density        & $k_{\rm max}$ & $\rm veh/m$ & 0.145  
\\
Capacity               & $c_l$      & $\rm veh/s$ & 0.51   
\\
Total lane length      & $D$      & $\rm m$     & 68631   
\\
Average lane length    & $l$      & $\rm m$     & 167    
\\
Average trip length    & $L$      & $\rm m$     & 7110    
\\
Signal cycle time      & $C$      & $\rm s$     & 50       
\\
Signal green time (mean)     & $\mu_G$      & $\rm s$     & 14.8    
\\
Signal green time (std.)     & $\sigma_G$   & $\rm s$     & 2.5     
\\
Offset                 & $\delta$ & $\rm s$     & 0            
\\ 
Traffic demand         & $m_0$      & $\rm veh/s$ & 0.6         
\\\bottomrule
\end{tabular}
\end{table}

\begin{figure}[hbt]
  \centering
    \subfigure[The MFD of the city center of Zurich through MoC]{%
      \resizebox*{8.3 cm}{!}{\includegraphics{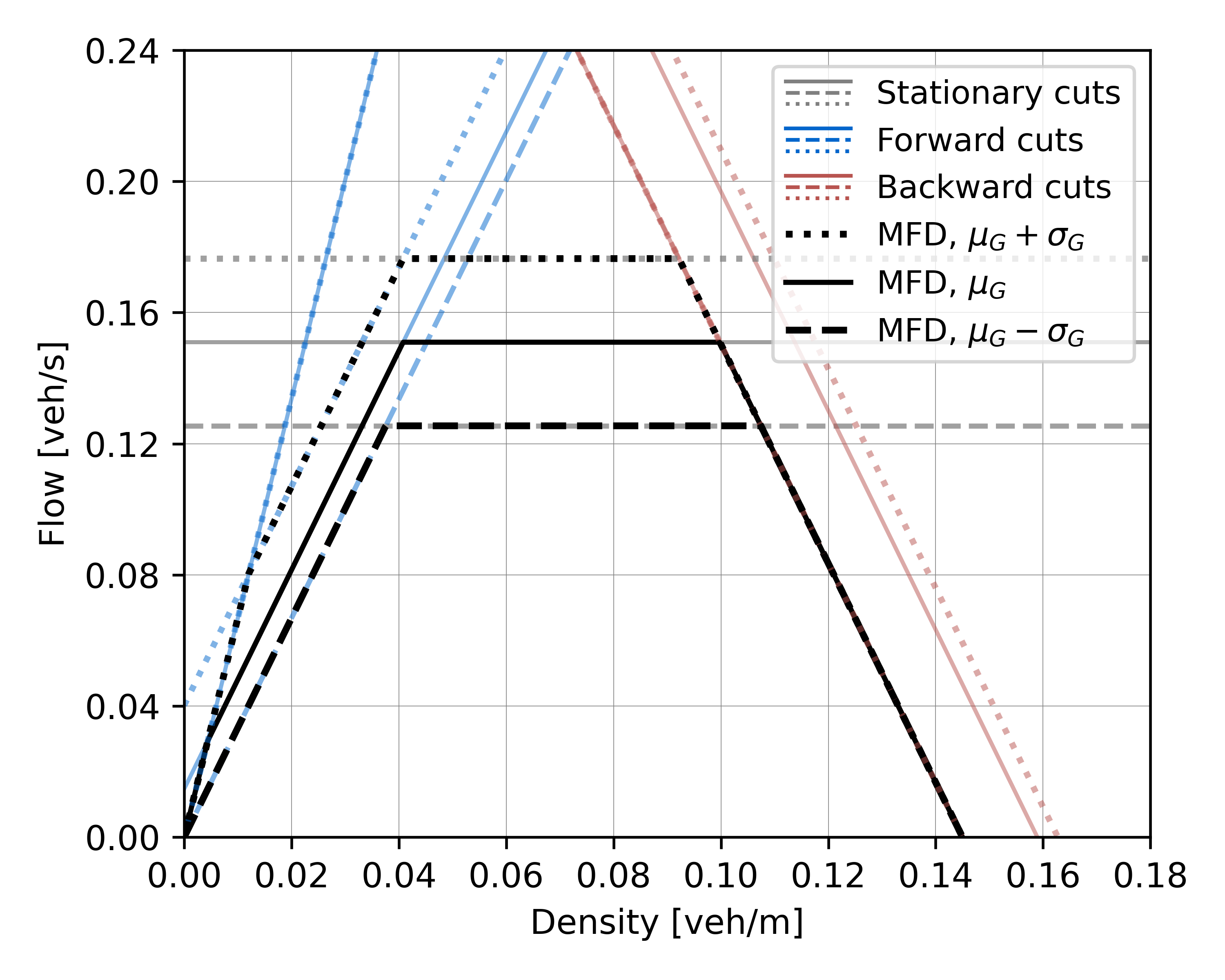}}
      \label{fig: zurich_mfd}}
    \subfigure[Stochasticity on the MFD]{%
      \resizebox*{8.3 cm}{!}{\includegraphics{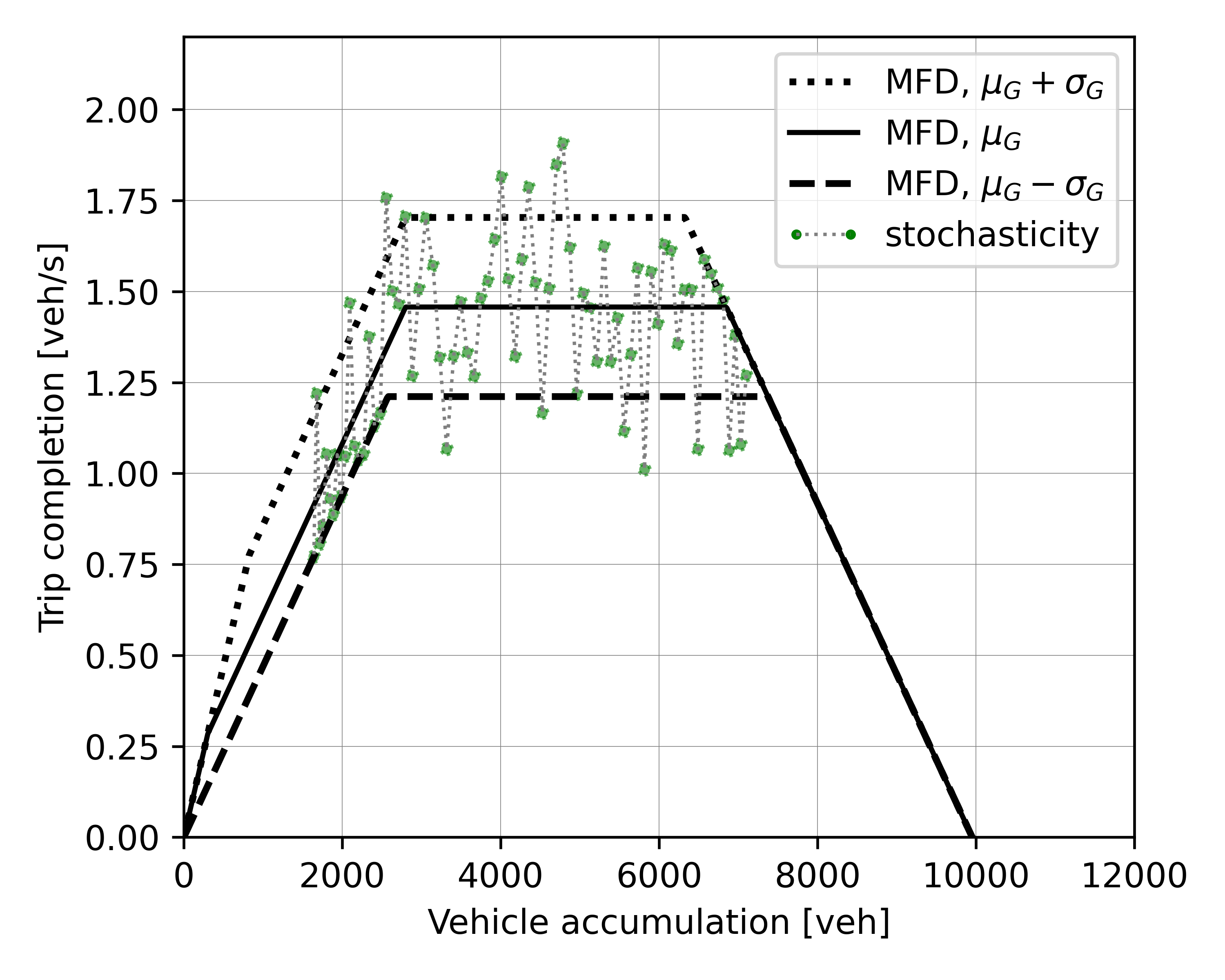}}
      \label{fig: stochasticity}}
    \caption{The MFD of the city center of Zurich through MoC and stochasticity.} 
\end{figure}

The average green time and the standard deviation of the signal are 14.8 s and 2.5 s, respectively. With an interval of one standard deviation, i.e., a set of green time $\mathcal{G}=\{\mu_G-\sigma_G, \mu_G, \mu_G+\sigma_G\}$, we can produce stationary, forward, and backward cuts in gray, blue, and red lines, for each of the MFD with a unique green time in $\mathcal{G}$ following MoC, as Fig. \ref{fig: zurich_mfd} shows with dotted lines, solid lines, and dashed lines. The MFD with a longer green time $\mu_G+\sigma_G$ yields a greater MFD and vice versa. The MFDs share the same maximal vehicle density of around 0.145 veh/m, which corresponds to a gridlock accumulation of about 10000 vehicles for the studied region based on Eq. \ref{eq: n} and Eq. \ref{eq: m}. In the following simulations for demand and supply disruptions with realistic model stochasticity, we study the congestion dissipation processes in the road network as traffic recovers from disruptions. At each timestep, a stochastic signal green time is sampled following a normal distribution $G\sim  \mathcal{N}(14.8,2.5)$ based on real-world data, leading to an uncertain MFD profile as Fig. \ref{fig: stochasticity} shows. The green scattering points connected with thin dotted lines display an illustration of a congestion dissipation process with an initial disruption of about 7000 vehicles and are sampled from every $100^{th}$ point for clarity. 

\subsection{Demand disruptions}

\begin{figure}[hbt]
  \centering
    \subfigure[Demand disruption and TTS]{%
      \resizebox*{8.3 cm}{!}{\includegraphics{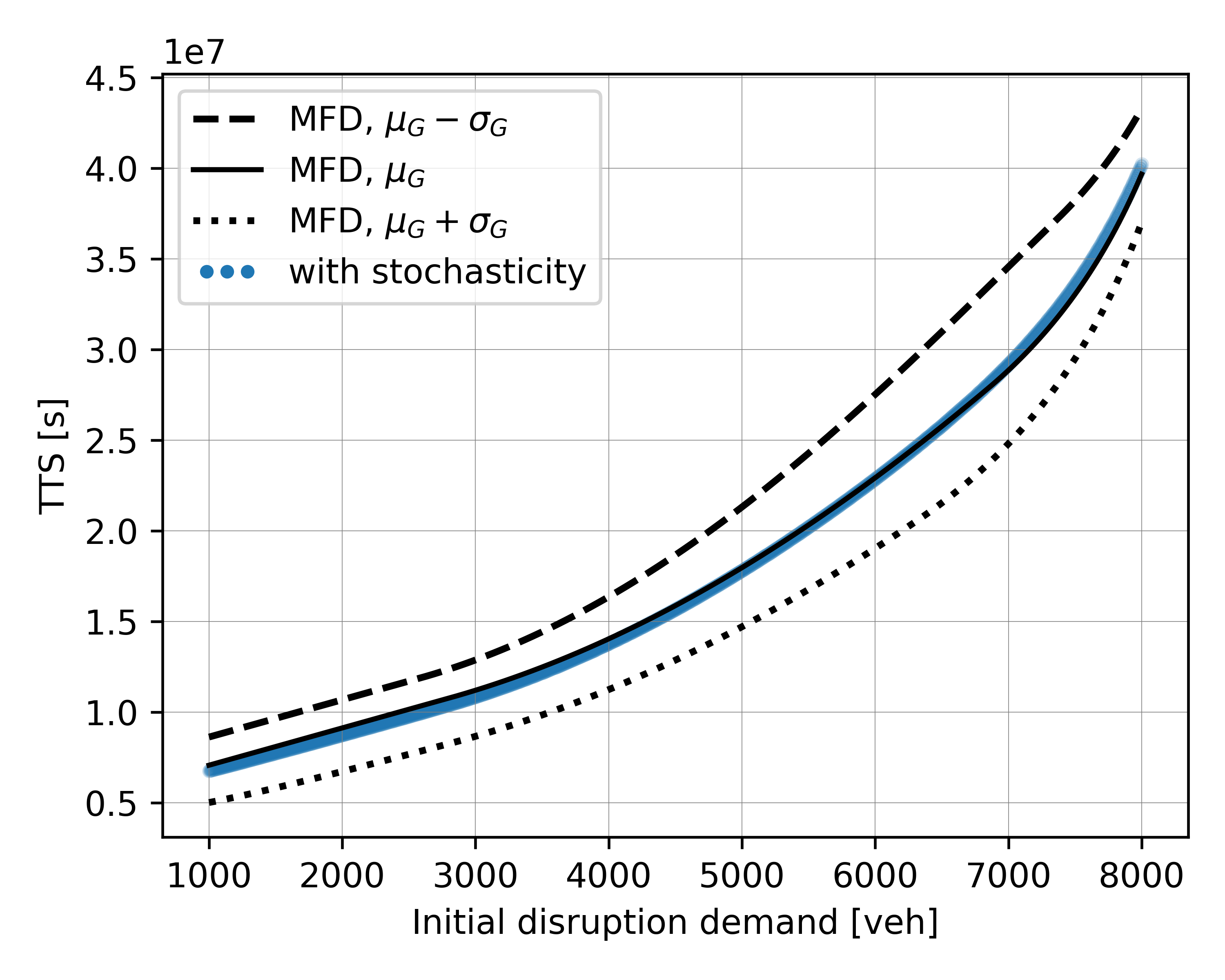}}
      \label{fig: TTS_demand}}
      \hspace{5pt}
    \subfigure[TTS distribution w/wo stochasticity ]{%
      \resizebox*{8 cm}{!}{\includegraphics{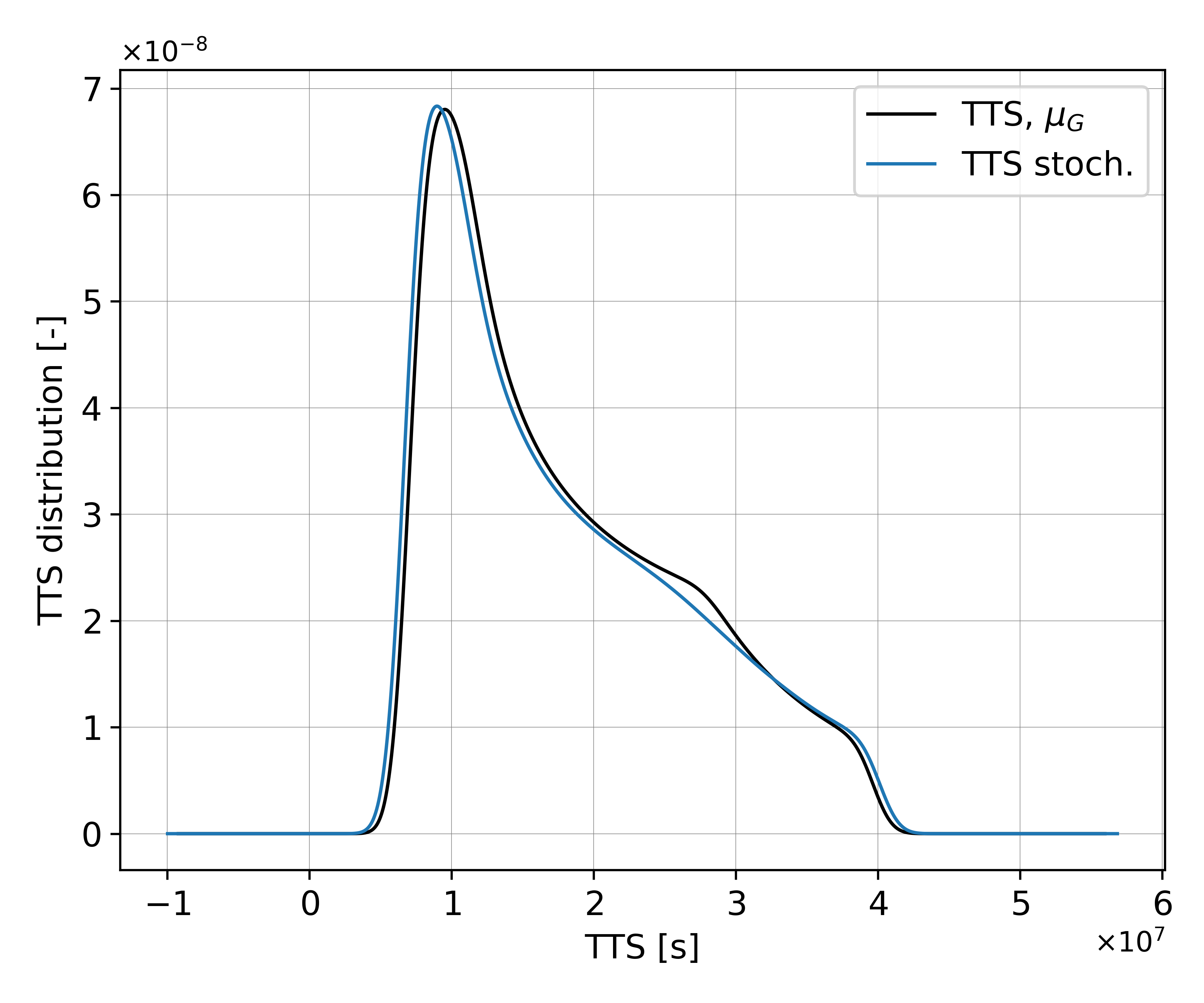}}
      \label{fig: distribution_demand}}
    \caption{Numerical simulation for demand disruption with stochasticity.} 
\end{figure}

Now we start the numerical simulation with different initial disruption demands $n'$ from 1000 to 8000 vehicles. The simulation time is 7200 seconds for each scenario with different initial demands. Fig. \ref{fig: TTS_demand} demonstrates that TTS grows exponentially with linearly increasing initial disruption demand, which validates the fragile nature proved with mathematical analysis. The solid, dashed, or dotted line each represents the TTS calculated under the three deterministic MFDs with green time in the set $\mathcal{G}$. Other than the black curves, there are also 1000 scattering points forming the blue curve. Each scatter point is composed of a full disruption recovery process sampled following a uniform distribution $n'\sim U(1000,8000)$. 

Since the blue curve, which is composed of the scattering points, closely aligns with the solid curve, it can be inferred that the influence of realistic stochasticity on the MFD is mostly negligible under demand disruptions. Nevertheless, an intriguing observation is that, when the disruption demand is relatively low, the blue curve dips slightly below the MFD of the solid curve. However, the curve appears to exceed the TTS of the well-defined MFD when the demand is substantial. This may indicate that the recovery process with stochasticity can have a larger second derivative. When we show the distribution of these two curves, as in Fig. \ref{fig: distribution_demand}, the TTS with stochasticity has a more concentrated distribution at a lower value while having a marginally longer tail pointing to the right, showing a more right-skewed distribution compared to the one without realistic stochasticity. This can also be validated by calculating the skewness of these two curves, with the skewness being 0.67 when there is no stochasticity, while the skewness for the blue curve has a value of 0.70. As a greater skewness indicates a more fragile network, it means that after introducing realistic stochasticity, the urban road network becomes even more fragile under demand disruptions. It makes particular sense that, as per the definition, a fragile network should exhibit a more degraded performance with larger disruptions brought by stochasticity, resulting in poor adaptability to uncertainties. 

\subsection{Supply disruptions}

\begin{figure}[hbt]
  \centering
    \subfigure[Supply disruption and the TTS]{%
      \resizebox*{8.3 cm}{!}{\includegraphics{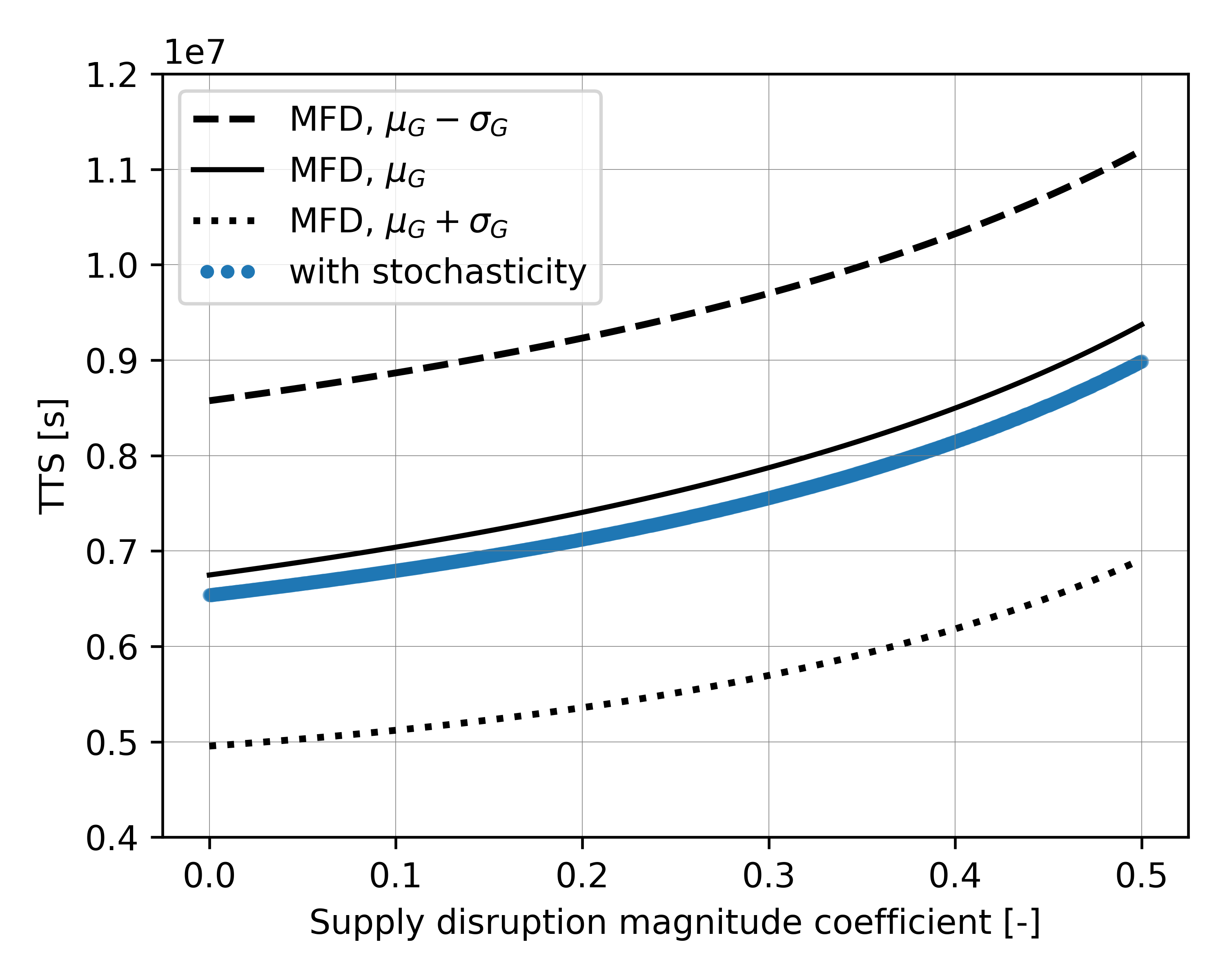}}
      \label{fig: TTS_supply}}
      \hspace{5pt}
    \subfigure[TTS distribution w/wo stochasticity]{%
      \resizebox*{8 cm}{!}{\includegraphics{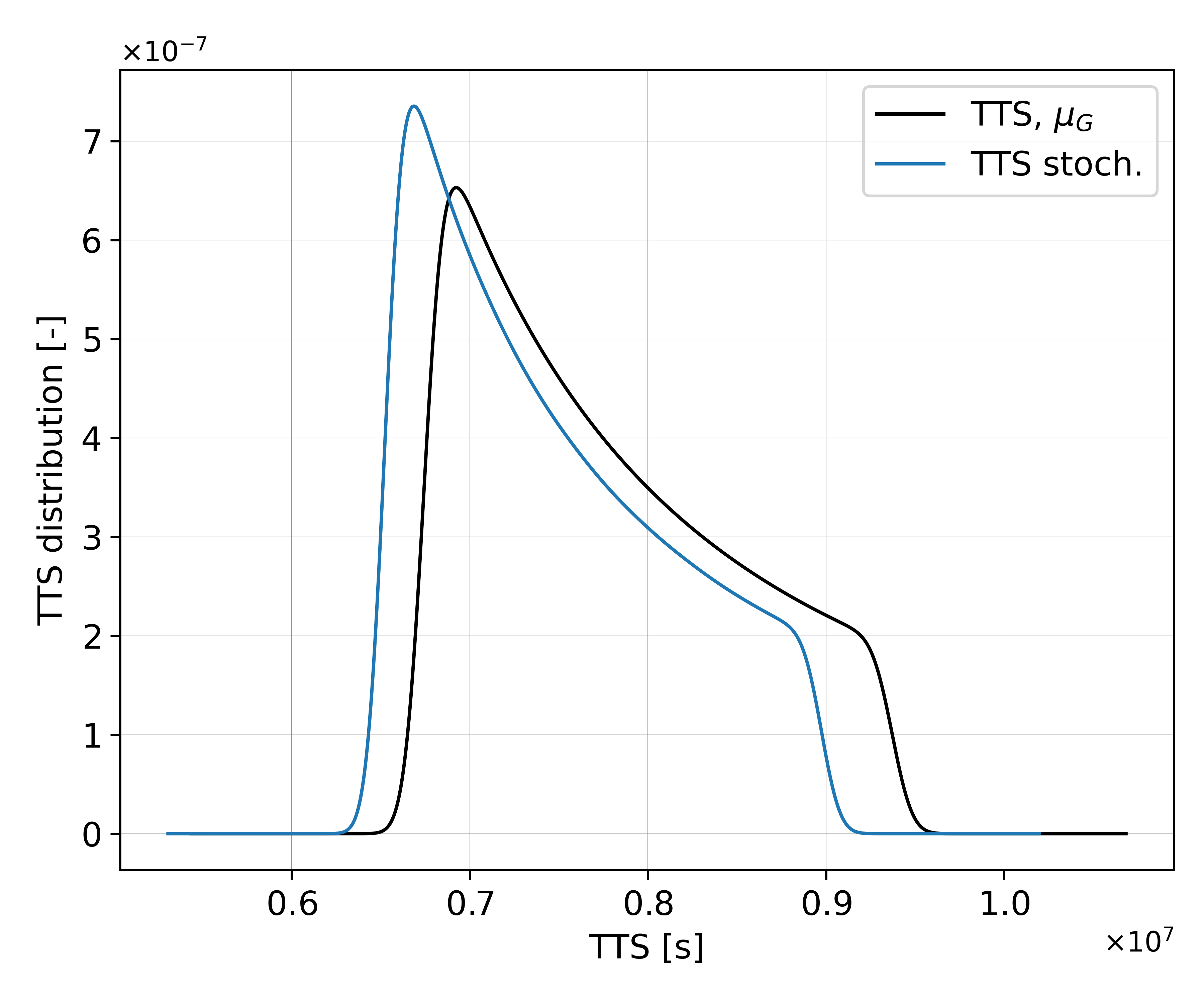}}
      \label{fig: distribution_supply}}
    \caption{Numerical simulation for supply disruption with stochasticity.} 
\end{figure}

Likewise, we showcase that supply disruptions can strengthen the fragile nature of transportation networks. With the same simulation environment, instead of linearly increasing initial disruption demand, a linearly growing supply disruption magnitude coefficient $r$ is implemented, growing from 0 to 0.5. The simulation of the recovery process from the supply disruption, along with 1000 uniformly sampled points $r\sim U(0,0.5)$, is shown in Fig. \ref{fig: TTS_supply}. As the blue curve with stochasticity lies below the curve from the deterministic MFD with the mean green time $\mu_G$, it means that the network with stochasticity has a better performance, which can be attributed to the uneven distribution of the trip completion rate due to signalization stochasticity in the set $\mathcal{G}$ on the free-flow cut. Despite this gain in the system performance, the distribution skewness is computed to be 0.49 for the deterministic MFD and 0.53 for the MFD with uncertainty, demonstrating that the network is more fragile under supply disruptions. 

Based on the results of the numerical simulations, this work confirms that stochasticity escalates the fragile response of road transportation networks under both demand and supply disruptions. 

\section{Conclusion}
\label{sec: conclusion}

This research introduces the pioneering concept of (anti-)fragility and its detection in the context of transportation. Then it systematically demonstrates the fragile nature of road transportation networks through rigorous mathematical analysis. With $m-n$ MFD to determine the system dynamics for a network, the second derivative of the performance loss over the magnitudes of disruptions can be proven to be positive, indicating the fragile property. Such fragility is validated under both demand and supply disruptions. Furthermore, this research also proposes a generic approach for quantifying network fragility using a scalable unit MFD and a skewness-based indicator. An approximation function inspired by the sigmoid function is developed to compute the skewness with high accuracy, enabling the cross-comparison of such fragile properties of different networks using merely MFD-related parameters with physical meanings. The proposed indicator can also be applied to evaluate the fragility of future infrastructures and influence policy-related decision-making. The indicator further provides insights into the potential benefits of speed limits and AVs in rendering networks less fragile. Additionally, through a numerical simulation with real-world data, including topology attributes, driving behavior, and signalization, among others, results suggest that stochasticity in the real world has a limited but reinforcing effect on the fragile characteristics of road transportation networks. 

Several limitations of this study need to be acknowledged. Although uncertainties are explicitly accounted for in the numerical simulation, this research largely relies on the assumptions of homogenous networks and well-defined MFDs. Since hysteresis is a commonly observed phenomenon on the MFDs due to unevenly distributed traffic congestion and unadaptive driving behavior, it can be of great interest to explicitly involve heterogeneity and hysteresis in future studies of antifragility. Proposition \ref{prop: 1} establishes that the recovery process can be considered fragile as long as the recovery branch of the MFD consists of multiple cuts forming a concave envelope, and various studies have also showcased that such concavity is still followed by hysteresis \citep{geroliminis_hysteresis_2011, mariotte_macroscopic_2017}. However, to fully incorporate hysteresis into the mathematical analysis on fragility, an analytical framework for the level of hysteresis must be present, which is sometimes referred to as the network stability \citep{gayah_impacts_2014}. Another limitation is the assumption that the backward wave is slower than the free flow speed in Section \ref{sec: indicator}, which allows for a fairly accurate skewness estimation using the proposed approximation function in this work. However, this assumption may be challenged by the emergence of AVs and platoons, highlighting the necessity for developing more precise approximation functions to address future technological advancements.

This study aims to offer insights to researchers, emphasizing the fragile nature of road transportation networks. Potential extensions of this work can be multifold. First, this study lays a theoretical foundation for the design of antifragile traffic control strategies, countering the intrinsic fragility of road networks through induced antifragile properties, such as the learning-based antifragile perimeter control in \cite{sun_antifragile_2024} or nonlinear variable structure control strategies in \cite{axenie2022antifragile}. Furthermore, as early findings have pointed out the possibilities of applying MFDs in other transportation modes, such as in railway \citep{saidi_train_2023}, the fragility demonstrated in road transportation networks may well be extended to various transportation systems or even systems in other disciplines with similar response. Given that network performance assessment in modern times should not rely on a sole criterion based on efficiency, a multi-objective framework can also be developed in the future, incorporating factors such as efficiency, antifragility, sustainability, safety, and more.

\section{CRediT authorship contribution statement}
Linghang Sun: Conceptualization, Investigation, Methodology, Visualization, Writing – original draft. Yifan Zhang: Methodology, Visualization, Writing - review \& editing. Cristian Axenie: Conceptualization, Project administration, Resources, Writing - review \& editing. Margherita Grossi: Project administration, Resources, Writing - review \& editing. Anastasios Kouvelas: Supervision, Writing - review \& editing. Michail A. Makridis: Conceptualization, Methodology, Supervision, Writing - review \& editing.

\section{Acknowledgement}
This research was primarily funded by the Huawei Munich Research Center through the Antigones project. Additional financial support has been received from the Swiss State Secretariat for Education, Research, and Innovation (SERI) with contract number 25.00120, as part of the  European Union's Horizon Europe project AntifragiCity with project ID 101203052. The authors would also like to thank the Reviewers for their constructive comments and Ying-Chuan Ni for the fruitful discussions.

\section{Declaration of Competing Interest}
This research was kindly funded by the Huawei Munich Research Center under the framework of the Antigones project, with one of our co-authors being employed at the said company. Otherwise, the authors declare that they have no known competing financial interests or personal relationships that could have appeared to influence the work reported in this paper.

\appendix
\section{Scaling a random MFD to the unit MFD}
\label{sec: scale}

\begin{figure}[hbt!]
\includegraphics[width=0.6\textwidth]{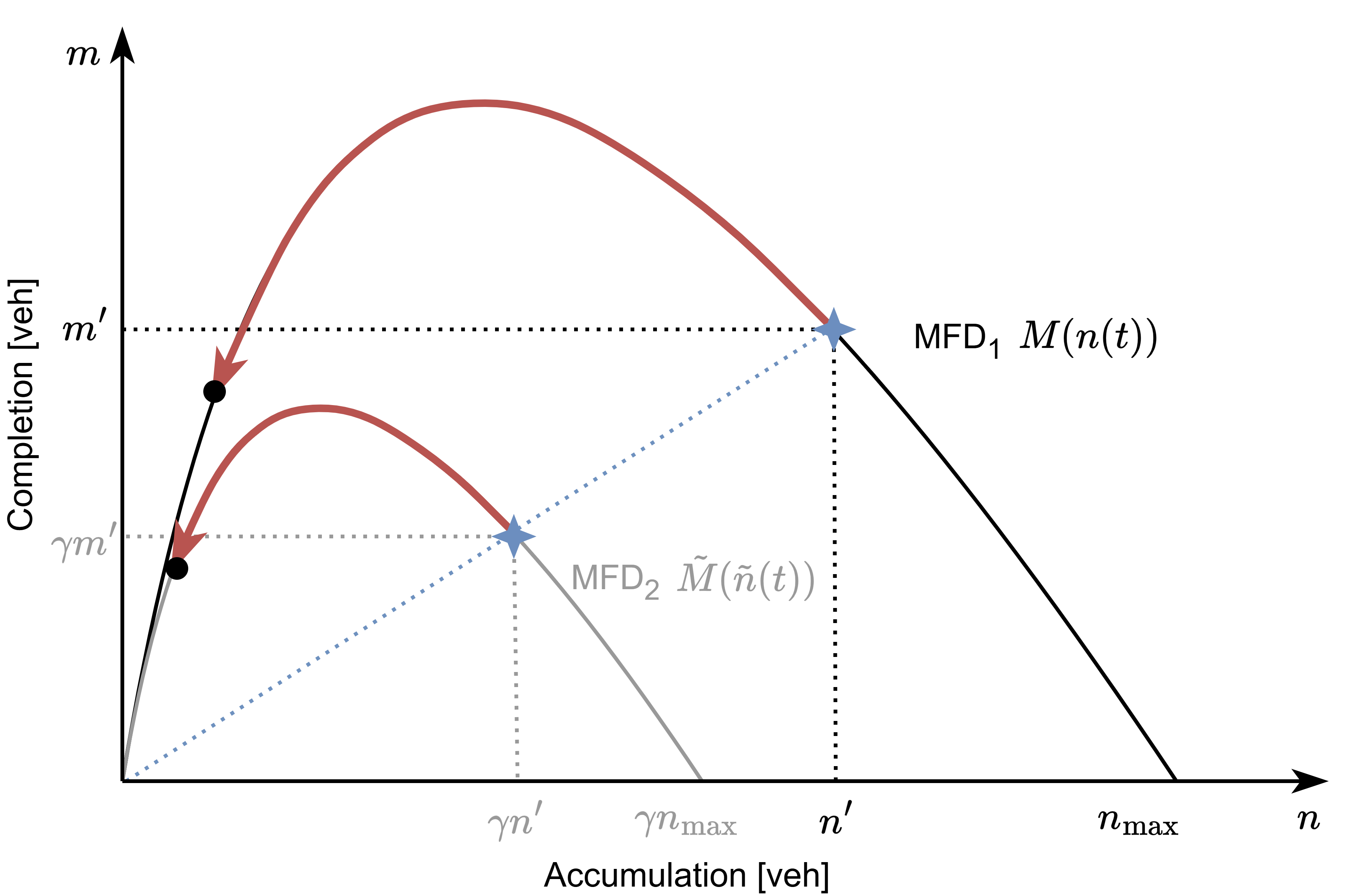}
\centering
\caption{Traffic state recovering from a supply disruption.}
\label{fig: scale}
\end{figure}

To prove a random MFD can be scaled to the unit MFD while preserving a fair comparison of their fragile properties, we need to verify that the skewness calculated based on Eq. \ref{eq: skewness} remains constant after scaling. As illustrated in Fig. \ref{fig: scale}, a generic $m-n$ MFD is presented alongside its scaled counterpart representing the unit MFD, which is adjusted using a scaling factor $\gamma$. As we sample initial disruption demand based on the percentage of maximal accumulation, the disruptive demand on the scaled MFD should also be scaled as $\gamma n'$ as well as the scaled maximal accumulation $\gamma n_{\rm max}$.

Since we assume no base demand when designing the fragility indicator, similar to Eq. \ref{eq: dynamics}, the system dynamics and the TTS for this original MFD are:

\begin{subequations}
\begin{alignat}{2}
\label{eq: dynamics_orig}
\frac{dn(\tau)}{d\tau} &= -M(n(\tau))
\\
\label{eq: TTS_original}
TTS_{orig} &= \int^t_0 n(\tau) d\tau
\end{alignat}
\end{subequations}

Likewise, with $\tilde{M}(\cdot)$ and $\tilde{n}$ each denoting the unit MFD and the vehicle accumulation, the system dynamics after scaling will be:

\begin{equation}
\label{eq: dynamics_unit}
\frac{d\tilde{n}(\tau)}{d\tau} = \tilde{M}(\tilde{n}(\tau)) = -\gamma M\left(\frac{\tilde{n}(\tau)}{\gamma}\right)
\end{equation}

We substitute $u(\tau)=\frac{\tilde{n}(\tau)}{\gamma}$ so that $\frac{d\tilde{n}(\tau)}{d\tau}=\gamma \frac{du(\tau)}{d\tau}$, and Eq. \ref{eq: dynamics_unit} can be rewritten as:

\begin{subequations}
\begin{alignat}{2}
\gamma\frac{du(\tau)}{d\tau} &= -\gamma M\left( u(\tau) \right)
\\
\Longrightarrow \frac{du(\tau)}{d\tau} &= - M\left( u(\tau) \right)
\end{alignat}
\end{subequations}

It shows that both $n(\tau)$ and $u(\tau)$ are governed by the same function while $\tilde{n}(\tau)=\gamma u(\tau)$. And the TTS on the unit MFD after scaling and its relationship with the TTS of the original MFD is:

\begin{subequations}
\begin{gather}
\label{eq: TTS_convert}
TTS_{unit} = \int^t_0 \tilde{n}(\tau) d\tau = \int^t_0 \gamma u(\tau) d\tau = \gamma \int^t_0 n(\tau) d\tau = \gamma TTS_{orig}
\end{gather}
\end{subequations}

To compute the skewness-based fragility indicator of the network with the original MFD using Eq. \ref{eq: skewness}, a list of $TTS_{orig,i}$ with $i\in\{1,2,\cdots,N\}$ should be generated based on a list of disruption demand $n'_{i}$. Following the above analysis, we obtain a list of $TTS_{unit,i}$ after scaling the original MFD to the unit MFD with the scaling factor $\gamma$. It can be easily proven that the mean $\mu$ and standard deviation $\sigma$ of the sampled TTS follow the same scaling factor $\gamma$. So the distribution skewness on the unit MFD is:

\begin{subequations}
\begin{alignat}{2}
s_{unit} = \frac{1}{N}\sum_{i=1}^{N}\left(\frac{TTS_{unit,i}-\mu_{unit}}{\sigma_{unit}}\right)^3 = \frac{1}{N}\sum_{i=1}^{N}\left(\frac{\gamma TTS_{orig,i}-\gamma \mu_{orig}}{\gamma \sigma_{orig}}\right)^3 = s_{orig}
\end{alignat}
\end{subequations}

We demonstrated that scaling a random MFD to the unit MFD does not alter its skewness.

\bibliographystyle{elsarticle-harv} 
\bibliography{references}
\end{document}